\RequirePackage[bookmarksnumbered,unicode]{hyperref}
\documentclass[acmsmall, screen, nonacm]{acmart}

\AtBeginDocument{%
  }

\setcopyright{acmcopyright}
\copyrightyear{2018}
\acmYear{2018}
\acmDOI{XXXXXXX.XXXXXXX}

\acmConference[Conference acronym 'XX]{Make sure to enter the correct
  conference title from your rights confirmation emai}{June 03--05,
  2018}{Woodstock, NY}
\acmPrice{15.00}
\acmISBN{978-1-4503-XXXX-X/18/06}

\usepackage{multirow}
\usepackage{makecell}
\usepackage{xspace}
\usepackage{mathtools}

\usepackage{tikz,pgfplots,graphicx,xcolor}
\pgfplotsset{compat=1.16}
\usepackage{wrapfig}
\usepackage[normalem]{ulem}

\usepackage{multirow}
\usepackage{tabularx}
\usepackage{array}

\usepackage{listings}
\usepackage{hyperref}
\usepackage[ruled,linesnumbered,algo2e]{algorithm2e}

\newcommand{\revision}[1]{{#1}}
\newcommand{\revisiontwo}[1]{{#1}}

\newcommand{\mypara}[1]{\smallskip\noindent\emph{\textbf{#1}.\ }}

\newcommand{\vmtl}{VMTLC\xspace}
\newcommand{\vmtlmod}{VMTLC-Mod\xspace}

\newcommand{\Ss}{\mathcal{S}}
\newcommand{\Rr}{\mathcal{R}}
\newcommand{\Ff}{\mathcal{F}}

\newcommand{\Ll}{\mathcal{L}}
\newcommand{\Cc}{\mathcal{C}}
\newcommand{\Ii}{\mathcal{I}}
\newcommand{\Aa}{\mathcal{A}}
\newcommand{\MM}{\mathit{Sol}}
\newcommand{\Ex}{\mathcal{E}}
\newcommand{\Tok}{\ensuremath{\mathit{TestOk}}}
\newcommand{\CHCok}{\ensuremath{\mathit{CHCOk}}}
\newcommand{\Pex}{\ensuremath{\mathit{PosCex}}}
\newcommand{\rel}{\ensuremath{r}}

\newcommand{\vx}{\ensuremath{\vec{x}}}
\newcommand{\such}{.\,}

\newcommand{\testres}{\ensuremath{\mathit{RB}}}
\newcommand{\tuple}[1]{\langle #1 \rangle}
\newcommand{\testprog}{\ensuremath{\mathit{TestProgram}}}

\newcommand{\inputbuf}{\ensuremath{\mathit{Buf}}}

\newcommand{\clientprog}{\ensuremath{C}}
\newcommand{\lib}{\ensuremath{L}}
\newcommand{\modmode}{\ensuremath{\mathit{modular}}}
\newcommand{\conmode}{\ensuremath{\mathit{contextual}}}
\newcommand{\recval}{\ensuremath{\mathit{val}}}

\newcommand{\vmtlc}{\textsc{Dualis}\xspace}
\newcommand{\hice}{\textsc{HornICE}\xspace}

\newcommand{\llm}{\textsc{LLM}\xspace}
\newcommand{\hllm}{\textsc{HornICE+LLM}\xspace}
\newcommand{\zt}{\textsc{Z3}\xspace}

\newcommand{\afl}{AFL++\xspace}

\newcommand{\true}{\ensuremath{\mathit{true}}}
\newcommand{\false}{\ensuremath{\mathit{false}}}
\newcommand {\union} {\cup}

\newcommand {\myOr} {\ensuremath{\vee}}
\newcommand {\myAnd} {\ensuremath{\wedge}}

\newcommand {\myImplies} {\ensuremath{\Rightarrow}}

\newcommand{\varset} {\ensuremath{\mathit{S}}}

\newcommand{\fnbstinsert}{\ensuremath{\texttt{insert}}}
\newcommand{\fnbstsearch}{\ensuremath{\texttt{search}}}

\newcommand{\varbstempty}{\ensuremath{\mathit{empty}}}
\newcommand{\varbstmin}{\ensuremath{\mathit{min}}}
\newcommand{\conbstinit}{\ensuremath{\mathit{init}}}
\newcommand{\conbstinsert}{\ensuremath{\mathit{insert}}}
\newcommand{\conbstsearch}{\ensuremath{\mathit{search}}}
\newcommand{\invbst}{\ensuremath{\mathit{inv}}}
\newcommand{\invoneset}{\ensuremath{\mathit{inv}_1}}
\newcommand{\invtwoset}{\ensuremath{\mathit{inv}_2}}

\newcommand{\fnsetinit}{\ensuremath{\texttt{init}}}
\newcommand{\fnsetempty}{\ensuremath{\texttt{empty}}}
\newcommand{\fnsetsize}{\ensuremath{\texttt{size}}}
\newcommand{\fnsetmin}{\ensuremath{\texttt{min}}}
\newcommand{\fnsetinsert}{\ensuremath{\texttt{insert}}}
\newcommand{\fnsetremove}{\ensuremath{\texttt{remove}}}
\newcommand{\varsetempty}{\ensuremath{\mathit{empty}}}
\newcommand{\varsetsize}{\ensuremath{\mathit{size}}}
\newcommand{\varsetmin}{\ensuremath{\mathit{min}}}
\newcommand{\varsetsum}{\ensuremath{\mathit{sum}}}
\newcommand{\consetinsert}{\ensuremath{\mathit{insert}}}
\newcommand{\consetremove}{\ensuremath{\mathit{remove}}}
\newcommand{\consetempty}{\ensuremath{\mathit{empty}}}
\newcommand{\consetinit}{\ensuremath{\mathit{init}}}

\newcommand{\this}{\ensuremath{\mathit{o}}}

\newcommand{\algmain}{%
    
\begin{algorithm2e}
  \SetAlgoSkip{}
  \SetAlgoNoLine
  \SetKwFor{For}{for}{do}{}
  \SetKwFor{While}{while}{do}{}
  \SetNoFillComment
  \KwIn{Client program $\clientprog$, library $\lib$, Mode $\mathit{mode}$
    ($\modmode/\conmode$), CHCs $\varphi$ 
    for $\clientprog$ over relations $\Rr = \Ii \cup \Aa$, 
    test resources~$\testres$ } 
  \KwOut{Solution $\MM$ to $\varphi$, in the form of a map from $\Rr$
    to predicates } 
  \BlankLine
  $\Pex \gets \emptyset$, $\Tok \gets \false$, $\CHCok \gets
     \false$\;\label{alg:main:init} 
  \While{$\neg (\CHCok \land \Tok)$}{
    $\tuple{\MM, \CHCok} \gets \textsc{ICE-CHCSolver}(\varphi,
      \Pex)$\;\label{alg:main:gcs}
    \uIf{$\mathit{mode}=\conmode$}{
      $\tuple{\Tok, P} \gets \textsc{TestGenerator}(\clientprog, \lib,
      \conmode, -, \MM, \testres)$\;\label{alg:main:tc}
      \lIf{$\neg \Tok$} {
        $\Pex \gets \Pex \union P$\label{alg:main:pc}
      }
    }
    \uElse(\tcp*[h]{$\modmode$ mode, test each library method $N$ in $\L$}){
      $\Tok \gets \true$ \;
      \ForEach{ method $N$ in $L$ }{
          $\tuple{\Tok', P} \gets$ \textsc{TestGenerator}($\clientprog,
             \lib, \modmode, N, \MM, \testres)$\;\label{alg:main:tm}
          \lIf{$\neg\Tok'$} {
            $\Pex \gets \Pex \cup P$ \label{alg:main:pm}
          }
          $\Tok \gets \Tok \land \Tok'$ \;
       }
    }
}
\Return{$\MM$}\;\label{alg:main:r}
\caption{\textsc{VMTLC}($\clientprog$, $\lib$, $\mathit{mode}$, $\varphi$, $\testres$)
\revisiontwo{}}
\label{alg:main}
\end{algorithm2e}

}

\newcommand{\alghice}{%
\begin{algorithm2e}
  \SetAlgoSkip{}
  \SetAlgoNoLine
  \SetKwFor{For}{for}{do}{}
  \SetKwFor{While}{while}{do}{}
  \SetNoFillComment
  \KwIn{$\varphi$ -- system of CHCs for a client program over relations $\Rr = \Ii \cup \Aa$, where $\Ii$ and $\Aa$ represent program invariants and library method contracts respectively; $\Pex$ -- set of positive examples for relations in $\Aa$ }
  \KwOut{$\MM$ (map from $\Rr$ to predicates) which is a solution to $\varphi$}
  \BlankLine
  $\Ex \gets \Pex$, $\MM_{cand} \gets \{\rel \mapsto \true\,|\, \rel \in \Rr \}$\;\label{alg:hice:init}
  \While{exists $\varphi_C$ such that $\varphi_C(\MM_{cand})$ is not valid\label{alg:hice:ex}}{
  $\Ex \gets \Ex \cup \textsc{GetCex}(\varphi_C, \MM_{cand})$\;\label{alg:hice:cex}
  $\MM_{cand} \gets \textsc{ICELearner}(\Ex)$\;\label{alg:hice:dt}
  }
\revisiontwo{\Return{$\tuple{\MM_{cand}, true}$}}\;\label{alg:hice:r}
\caption{\textsc{ICE-CHCSolver}($\varphi$, $\Pex$)}
\label{alg:hice}
\end{algorithm2e}
}

\newcommand{\algtest}{%
\begin{algorithm2e}
  \SetAlgoSkip{}
  \SetAlgoNoLine
  \SetKwFor{For}{for}{do}{}
  \SetKwFor{While}{while}{do}{}
  \SetNoFillComment
  \KwIn{Client program $\clientprog$, library $L$, method under test
    $N$, contracts $\MM$ for methods in $L$, resource bound on testing
    $\testres$} 
  \KwOut{Result of testing $\Tok$, set of positive examples $P$}
  \BlankLine
  $\testprog \gets
  \textsc{ConstructProgram}(\clientprog, \lib, \mathit{mode}, N, \MM)$\; \label{alg:test:add} 
  \While{$\testres$ is not exhausted\label{alg:test:testbound}}{
  $\inputbuf \gets \textsc{InitBuf()}$\;\label{alg:test:init}
  $\tuple{\Tok,N',\recval} \gets \textsc{Execute}(\testprog, \inputbuf)$\;\label{alg:test:exec}
  \lIf{$\neg \Tok$}{\Return $\tuple{\false, \{(N',\recval)\}}$}\label{alg:test:retfail}
  }
  \Return $\tuple{\true, \emptyset}$\;\label{alg:test:retsuc}
  \caption{\textsc{TestGenerator}($\clientprog$, $\lib$,
    $\mathit{mode}$, $N$, $\MM$, $\testres$) \revisiontwo{}}
\label{alg:test}
\end{algorithm2e}

}

\begin{document}

\title{Verification Modulo Tested Library Contracts}


\author{Abhishek Uppar}
\orcid{0000-0002-0081-6945}
\email{abhisheku@iisc.ac.in}
\affiliation{%
  \institution{Indian Institute of Science}
  \city{Bangalore}
  \state{Karnataka}
  \country{India}
}

\author{Omar Muhammad}
\orcid{0009-0008-2086-809X}
\email{omarmuhammad@iisc.ac.in}
\affiliation{%
  \institution{Indian Institute of Science}
  \city{Bangalore}
  \state{Karnataka}
  \country{India}
}
\author{Sumanth Prabhu S}
\orcid{0009-0009-1105-5529}
\email{sumanthsprabhu@gmail.com}
\affiliation{%
  \institution{Relyance AI}
  \city{Bangalore}
  \state{Karnataka}
  \country{India}
}

\author{Deepak D'Souza}
\orcid{0000-0002-6629-6604}
\email{deepakd@iisc.ac.in}
\affiliation{%
  \institution{Indian Institute of Science}
  \city{Bangalore}
  \state{Karnataka}
  \country{India}
}

\author{P. Madhusudan}
\email{madhu@illinois.edu}
\orcid{0000-0002-9782-721X}
\affiliation{
  \institution{University of Illinois Urbana-Champaign, Department of Computer Science}
  \city{Urbana}
  \state{Illinois}
  \country{USA}
}

\author{Adithya Murali}
\orcid{0000-0002-6311-1467}
\email{adithya5@illinois.edu}
\affiliation{%
  \institution{University of Wisconsin}
  \city{Madison}
  \state{WI}
  \country{USA}
}


\begin{abstract}
We consider the problem of \emph{verification modulo tested library
  contracts} as a step towards automating the verification of client
programs that use complex libraries.
We formulate this problem as the synthesis of modular contracts for the
library methods used by the client that are adequate to prove the client correct, and
that also pass the scrutiny of a testing engine that tests the library against these contracts.
We also consider a new form of method contracts called \emph{contextual contracts} that arise in this setting
that hold in the context of the client program, and 
can often be simpler and easier to infer than classical modular contracts.
We provide a counterexample-guided learning framework to solve this
problem, in which the synthesizer interacts with a constraint solver
as well as the testing engine in order to infer adequate modular/contextual 
method contracts and inductive invariants for the client.
The main synthesis engines we use are generalizing CHC solvers that are realized using ICE learning algorithms.
We realize this framework in a tool called \vmtlc
and show its efficacy on benchmarks where clients
call large libraries.
\end{abstract}

\received{20 February 2007}
\received[revised]{12 March 2009}
\received[accepted]{5 June 2009}

\maketitle



\section{Introduction}
\label{sec:intro}

Formal verification mechanisms have made great strides in the last few decades, resulting in mature logics, verification paradigms, automation using logic engines, and tool frameworks. However, for expressive specifications of large programs, complete automation is not within reach. Effective verification methods for large software today call for human-written annotations, such as contracts, that enable modular and hence scalable verification. Furthermore, human annotations in terms of inductive invariants for iterative blocks of code are needed to enable scalable automation using logic solvers. While there are techniques for synthesizing inductive invariants and contracts~\cite{McMillan03,SomenziB11,houdini,spacer1,spacer2,ice}, inductive invariant synthesis is a hard search problem, and current synthesis techniques do not scale. We are at an impasse today as it has become clear that completely automated verification is unlikely to scale to large code and complex specifications. Automated verification, however, does work well for smaller pieces of code with complex specifications, or large code bases with simple specifications (e.g., null pointer dereference checking, data races, etc.). 

In this paper, we explore a more pragmatic choice in the design space---relaxing verification guarantees in order to achieve scalable and complete automation for larger code-bases.
In particular, we are interested in automatically verifying smaller \emph{client programs} that utilize large libraries. A traditional verification approach would require annotating the libraries with contracts strong enough to prove the client code, and then proving the client code and the libraries against their contracts using further annotations (such as inductive loop invariants).
Annotating and proving large libraries against their contracts is a daunting task; however, in many settings, the client code is smaller, and we can automate the synthesis of annotations  using current techniques.  

We negotiate this space by relaxing the guarantee of formal verification of libraries in order to scale verification. More precisely, our aim is to automatically synthesize contracts for libraries, \emph{formally} prove that the client is correct assuming these contracts, and, instead of formally verifying the library code, demand that the library pass extensive \emph{testing} against the contracts.

We propose the following novel \emph{verification modulo tested
  library contracts} problem:
Given a client program $C$ using a library $L$,  find a set of
\emph{formal contracts} $\Pi$ for the methods in $L$ such that (a) the
client $C$ is proved correct assuming the contracts $\Pi$ for $L$, and
(b) a given \emph{deterministic test-generator with a bound on its
  resources} is unable to find any tests for the library methods that
violate their contract in $\Pi$. 

Though the precise problem we define here seems new, there have been several variants of the above problem in recent years that have considered synthesizing contracts for libraries aided by testing only (and not verification)---for functional programs (albeit without synthesizing inductive invariants for clients)~\cite{ZhouDDJ21} and for 
synthesizing preconditions and contracts purely from libraries~\cite{astorgaprecond,astorgacontract} (see related work in Section~\ref{sec:relwork} for more details). 

In this paper, we seek entirely automated methods that solve the stated problem, both in terms of synthesizing contracts for libraries as well as synthesizing the inductive annotations for the client that prove it correct.
Furthermore, we are interested in the setting of imperative object-oriented programming languages, where we assume libraries implement classes that encapsulate data and the methods, where classes manage unbounded collections of objects (like container classes implemented using data structures), and classes and objects have \emph{state} that evolves with method calls. 

\mypara{Learning Library Contracts and Client Invariants Simultaneously}
The key problem to solve is one of synthesis---synthesizing contracts $\Pi$ for the libraries as well as synthesizing annotations for the client code (inductive loop invariants, contracts for client's methods, etc.) such that a logic engine can validate the verification conditions for the basic blocks in the client code, and at the same time ensure the library contracts are vetted correct by the given test generator.

The first main technical contribution of this paper is a learning
framework that simultaneously synthesizes adequate contracts for
libraries and inductive annotations for client code to solve the above
problem. We propose a \emph{counterexample guided learning/synthesis}
framework. In each round, we require the verifier to handle a set of
CHC (Constrained Horn Clause) constraints derived from the client,
$CHC(C)$, along with a set of \emph{positive samples} for library
methods. Intuitively, we require the verification engine to solve the
CHC constraints of the client in a way that the predicates associated with
library contracts in the CHC include the positive
samples. (These positive samples can also be encoded into CHC
constraints, of course; but the crucial aspect is that these positive
examples evolve during the verification process). In each round, we
require that the verification engine solves the CHC constraints
(thereby finding inductive annotations for the client and contracts
for the library methods) while ensuring that the
contracts for the library methods include the
positive samples.

It is important to our framework that the CHC solver \emph{generalize} as well (a generic CHC solver does not suffice). A CHC solver that overfits the contracts
(say allowing only the given positive examples to be the library's contract)
will \emph{not} be useful in our context--- the problem is not to solve the CHCs but to simultaneously generalize the library contracts and solve the CHCs.

Our framework pairs a generalizing solver for CHCs and positive
samples with a test generator, working in
rounds. In each round, we take the CHC solver proposed annotations,
and check if the testing engine can find violations of the library
against their synthesized contracts. If the testing engine is able to
produce failing tests, we extract positive examples that show
violation of the contract, augment the set of positive samples, and
iterate. If the testing engine is unable to falsify the contracts, we
can terminate, as we have verified the client modulo tested library
contracts.

Turning to realizing generalizing solvers for CHCs and positive examples, we propose ICE-learning engines~\cite{ice,icedt,DBLP:journals/pacmpl/EzudheenND0M18} as natural candidates. ICE-learning engines
strive to solve CHCs by learning inductive invariants and annotations in hypothesis classes, aiming to find simplest expressions that solve CHC constraints, and thus naturally generalize. Given CHC constraints, an ICE solver works itself in an iterative loop involving an ICE learner and a constraint solver. The ICE learner, in each iteration proposes inductive invariants that solve the CHCs, and the constraint solver responds using  \emph{positive, negative, and implication counterexamples} (ICE samples) when validity fails. 
Consequently, the positive examples generated by testing engines (in the outer loop) naturally fit into an ICE-learning framework as we can add them to the set of ICE samples the learner considers in each round.
%
The learning algorithm required in our framework can be any ICE-learning algorithm that synthesizes invariants and contracts given counterexamples. 
Moreover, when the hypotheses spaces of invariants and contracts is finite, our framework is guaranteed to converge and verify the client modulo tested library contracts, provided they exist.

\mypara{Contextual contracts}
The second contribution of this paper is the study
of a new form of contracts that emerges particularly in the setting of verification modulo tested libraries. Consider a client $C$ and a library $L$ that it utilizes. In traditional verification, contracts $\Pi$ for $L$ are valid if the methods in $L$ verify their contract when they are called in \emph{all possible states and all possible inputs}. However, since we are interested in verification primarily of the client, it is sufficient if $L$ satisfies the contracts $\Pi$, \emph{in the context of the client $C$}. More precisely, the library is not called in every possible state of object heap and input parameters---the client only calls it in particular states. Hence, there is an \emph{implicit precondition} that the overall program satisfies when calling library functions. Under these preconditions, the contracts for the libraries could be made stronger, which in turn can help prove the client correct. We call these contracts that hold in the context of the client \emph{contextual contracts}.

For example, consider a library that has a class that maintains a
datastructure for a set, with methods to insert and delete elements,
and a method to check membership. Now consider a client $C$ that only
inserts positive integers into a set maintained by such a datatype,
and then checks for membership of a negative number, and asserts that
this must return false. In this case, a contextual contract for the
membership method could state that it returns false for all negative
numbers. Note that this contract holds in the context of this client
as the client only adds positive integers to the container (and proves
the assertion in the client correct), but does not hold in
general.~\footnote{A contract that holds in all settings and proves
the client correct does exist but is much more complex---we would need preconditions that say that any number inserted into the set is positive, keep a class invariant that all elements stored in the set are positive (which will require quantification or recursive definitions).}

We propose the following verification modulo tested \emph{contextual} library contracts problem: given
a client program $C$ using a library $L$, find a set of formal contracts $\Pi$ for the library methods in $L$
at each call site in $C$ such that (a) the client $C$ is proved correct assuming the contracts $\Pi$ for $L$, and (b) a predetermined
test-generator with fixed resources is unable to find any tests for the program with the client and the library methods that violate
the contracts in $\Pi$.

Our framework can cater to synthesizing contextual contracts by testing the contracts \emph{only in the context of the client}. We modify our framework so that the testing phase is performed not modularly on the library, but on the entire program, involving the client and the library.

\mypara{Learning Algorithms: Search, Contraint-solving and LLMs}
As mentioned earlier, our framework allows the use of any ICE learning algorithm, and we use a decision-tree learning called HornICE~\cite{DBLP:journals/pacmpl/EzudheenND0M18}.

We can also use large language models (LLMs) in this context as ICE-learners. Note that we do not assume LLMs to produce reliable annotations as we will always validate them (using formal verification and testing). Given a client
program, and only a \emph{description} of the library, we can utilize
LLMs to propose contracts (and invariants for the client). When
proposals are found to be incorrect (either in verification of the
client or testing of libraries), we can feed the counterexamples back
to the LLM and ask it to modify its proposal, turning it into an ICE learner. While, of course, there are no guarantees that proposals would avoid the counterexamples or converge, the LLM can speculate on contracts using the documentation, code, and semantics of natural language. For example, an LLM easily predicts that isEmpty() should return \emph{false} after calling insert() in a container class, even with no access to library code. 

We also develop algorithms that combine LLMs with symbolic methods. In particular, we propose an ICE-learning algorithm that extracts predicates from the LLM proposed contracts to use in a symbolic learning algorithm to synthesize contracts that truly avoid counterexamples.

\revision{
\mypara{Limitations} Note that VMTLC with modular/contextual contracts can, of course, succeed on incorrect programs, as library contracts are not verified, but only tested. In order to succeed, spurious contracts would needed to be synthesized that pass testing, and inductive invariants for the client found that prove the client under these library contracts. While modular contracts are tested extensively in all possible contexts, contextual contracts are tested only in the context of the client, and hence are more susceptible to being spurious. See Section~\ref{subsec:VMTLCValue} for a  detailed discussion.
}

\mypara{Evaluation} 
To evaluate our framework, we build a tool called \vmtlc that implements our
framework for verifying clients modulo tested library contracts. Our
tool combines verification engines that return counterexamples, test
generators, and a variety of ICE learners. 
We evaluate our proposed techniques on a suite of benchmarks drawn from open-source data
structure libraries, most beyond the power of automated verification today. We evaluate (a) the efficacy of our
framework in solving the VMTLC problem, for several choices of generalizing CHC solvers, including LLMs (b) comparison of modular and contextual contracts, (c) whether synthesized contracts by the various approaches are indeed valid, and (d) ablations that study the effectiveness of generalizing CHC solvers and nongeneralizing ones. 

Our evaluation shows that our framework is effective on verifying clients modulo synthesized library contracts, and holds promise to scale verification.



\section{Illustrative Example}
\label{sec:illustratedeg}

In this section we illustrate the VMTLC problem via an example.
Consider the client program shown in
Fig.~\ref{fig:set-client}(a).
The program uses a Set library to maintain a set of integers.
It inserts a number of non-negative integers into the set, and then
removes them from the set while computing the sum of their values.
The programmer asserts that $\varsetsum$ variable contains a
non-negative value at the end of the program.
We would like to establish that the assertion is valid.

\begin{figure}
\begin{center}
\begin{minipage}[t]{0.45\textwidth}
\begin{lstlisting}[language=C++, basicstyle=\ttfamily, numbers=left,
    basicstyle=\small, morekeywords={assert,nondet, bool, assume, Set}]
int main() {
    Set S;
    int N = nondet();
    for (int i = 0; i < N; i++) {
        int v1 = nondet();
        if (v1 >= 0)
          S.insert(v1);
    }
    int sum = 0;
    while (!S.empty()) {
      int v2 = S.remove();
      sum = sum + v2;
    }
    assert(sum >= 0);
}
\end{lstlisting}
\end{minipage}
\ \ \ \ \ \ \ \ \ \ \ \ %
\begin{minipage}[t]{0.45\textwidth}
\begin{small}
\begin{lstlisting}[language=C++, basicstyle=\ttfamily,
    basicstyle=\small, morekeywords={assert,nondet, bool, assume, Set},mathescape]
void init(Set t); $\mbox{// Contract: } \revision{\varsetempty'(\this)}$
\end{lstlisting}
\vspace{-0.15cm}

\medskip
\begin{lstlisting}[language=C++, basicstyle=\ttfamily,
    basicstyle=\small, morekeywords={assert,nondet, bool, assume, BST,
      BinarySearchTree}]
void insert(Set t, int ival);
\end{lstlisting}
\vspace{-0.15cm}
// Contract: 
(($\varsetempty(\this) \myOr \varsetmin(\this) \geq 0) \myAnd \mathit{ival}
    \geq 0)) \myImplies \varsetmin'(\this) \geq 0$

\medskip
\begin{lstlisting}[language=C++, basicstyle=\ttfamily,
    basicstyle=\small, morekeywords={assert,nondet, bool, assume, Set}]
bool empty(Set t);
\end{lstlisting}
\vspace{-0.15cm}
// Contract: 
$\mathit{ret} \equiv \varsetempty(\this) \myAnd \varsetempty'(\this) \equiv
\varsetempty(\this)$

\medskip
\begin{lstlisting}[language=C++, basicstyle=\ttfamily,
    basicstyle=\small, morekeywords={assert,nondet, bool, assume, Set}]
int remove(Set t);
\end{lstlisting}
\vspace{-0.15cm}
// Contract:
$(\neg\varsetempty(\this) \myAnd \varsetmin(\this) \geq 0) \myImplies
(\mathit{ret} \geq 0) \myAnd (\varsetempty'(\this) \myOr
\varsetmin'(\this) \geq 0)$

\medskip
// Loop invariants: \\
$\mathit{inv1}:$ $\varsetempty(\varset) \myOr \varsetmin(\varset) \geq 0$ \\
$\mathit{inv2}:$ $\varsetsum \geq 0 \myAnd (\varsetempty(\varset) \myOr \varsetmin(\varset) \geq 0$) \\
\end{small}
\end{minipage} \\[-2mm]
\begin{minipage}[t]{0.4\textwidth}
\caption*{(a)}
\end{minipage}
\begin{minipage}[t]{0.45\textwidth}
\caption*{(b)}
\end{minipage}
\end{center}
\caption{(a) A client program that uses a \textsc{Set}
  library, and (b) Library methods along with contracts, and loop
  invariants, learnt by our approach.}
\label{fig:set-client}
\end{figure}

In addition to the $\fnsetinsert$, $\fnsetremove$, $\fnsetempty$ (and
the implicit $\fnsetinit$ constructor) methods,
let us say the Set library has a couple of ``observer'' methods
$\fnsetsize()$ and $\fnsetmin()$ that take no arguments and return
integer values. We will also consider the method $\fnsetempty()$ to be an
observer.
The observer methods $\fnsetsize$ and $\fnsetmin$ are meant to return
the size and the minimum value stored in the set, respectively, 
but our technique is agnostic to these semantics.
We associate uninterpreted functions $\varsetempty$ 
(of type $\mathrm{Set} \rightarrow \mathrm{Bool})$, and $\varsetsize$ and
$\varsetmin$ (of
type $\mathrm{Set} \rightarrow \mathrm{Int}$) with the observer methods,
to abstractly represent the state of the data structure maintained by
the library.

Our goal is now to come up with annotations for the client program
in the form of loop invariants
and method contracts, which constitute a valid Floyd-Hoare proof that
the program satisfies its assertion.
There are a couple of caveats though.
Firstly, the library method contracts must make use of only the
observer functions
(instead of the actual data structure state maintained by
the library), apart from the formal parameters and return values of
the method.
The method contracts may also refer to the auxiliary state
\emph{after} the 
invocation of the method, using primed versions of
the observer functions.
For example, the contract for $\fnbstinsert$ could be a predicate over
the variables $\this$ (representing the receiver object on which the
method is invoked), $\varsetempty(\this)$, $\varsetsize(\this)$,
$\varsetmin(\this)$,
$\varbstempty'(\this)$, $\varsetsize'(\this)$, and $\varbstmin'(\this)$
(representing the 
values returned by the observer functions in the state pre and post the method
invocation), and $\mathit{ival}$ (the formal parameter to $\fnsetinsert$).
The loop invariant $\mathit{inv1}$ for the for-loop would be a predicate over
the variables $N$, $i$ and
$\varset$ (variables in scope at the loop head), and \revision{likewise}
the loop invariant $\mathit{inv2}$ for the while-loop would be a
predicate over $N$, $\varset$, and $\varsetsum$.
%
The second caveat is that the method contracts
need only pass the given tester---they need not be formally proved.

For this program, the approach we describe in this paper essentially comes up
with the adequate testable annotations (library method contracts
and loop invariants) shown in Fig.~\ref{fig:set-client}(b).
Here $\mathit{ret}$ represents the return value of the methods
$\fnsetempty$ and $\fnsetremove$.
The annotations are adequate to prove the assertion in the client, as
well as pass the Tester.

\medskip
Finally, to illustrate the use of contextual contracts, consider the
same client program of Fig.~\ref{fig:set-client}(a), but where the library
does not provide the $\fnsetmin$ observer method.
%
%
There is clearly \emph{no} adequate testable modular contract for the library
methods using these observer methods.
However, what we \emph{can} say is that, in the \emph{context} of the given
client program,
the call to $\fnsetremove()$ at Line~11 satisfies the contract
``$\varsetempty(\this) \myOr \mathit{ret} \geq 0$''.
This contextual contract can be tested by the Tester by simply executing the
client program with different inputs and checking whether the call to
$\fnsetremove()$ at Line~11 satisfies this contract.
Our approach comes up with this contract and the other annotations below:
\[
\begin{array}{ll}
  \conbstinit \mbox{ (Line~2)} \mapsto \revision{\varsetempty'(\this)} &
    \consetempty \mbox{ (Line~10)} \mapsto
  \mathit{ret} \equiv \varsetempty(\this) \myAnd \varsetempty'(\this) \equiv
  \varsetempty(\this), \\
  \consetinsert \mbox{ (Line~7)} \mapsto \true & \consetremove \mbox{ (Line~11)} \mapsto \varsetempty(\this) \myOr \mathit{ret} \geq
  0 \\
  \mathit{inv1} \mapsto \true & \mathit{inv2} \mapsto \mathit{sum}
  \geq 0
\end{array}
\]
which turns out to be adequate to prove the assertion in the
client, and also pass the contextual Tester.

\section{The Verification Modulo Tested Library Contracts Problem}
\label{sec:prelim}

In this section we formally define the main technical objects of study, namely the verification modulo tested library contracts (\vmtl) problem and the concept of contextual contracts. We begin with some preliminaries to develop the vocabulary for our definitions.

\smallskip
\mypara{Classes and Objects} In this work we study the \vmtl problem in the domain of object-oriented programs. We work with a client program $C$ (with a \textit{main} method) that creates and manipulates objects of classes that are part of a library. Let us consider a set of primitive sorts (mainly integers and Booleans in this paper).
We model a class $c$ using a designated sort $O_c$ for objects of that class. 
We will use these to explicitly model manipulation of objects 
by methods. A class $c$ consists of a set of methods $\{f_\textit{cons}(\overline{p_\textit{cons}}), 
f_1(o,\overline{p_1}), \ldots, f_n(o,\overline{p_n})\}$, 
where $f_\textit{cons}$ is a constructor method to create objects, and $f_1,\ldots, f_n$ are methods on receiver object $o$ given as the first parameter (a method call  \texttt{o.f(p)} is modeled as calling $f(o,p)$).
These methods in general have side effects, and return a reference to an object in $O_c$ as well as a return value in a primitive sort. We don't formalize programming languages further as our  problems are meant to work on real programming languages; we assume that they come with a sound verification condition generator for verifying programs with adequate annotations, and a test generator for testing programs against annotations. 

\mypara{Observer Methods and Annotations} We now define the syntax and semantics for annotations (both invariants and contracts) that we consider in this work. 

Let us fix a  first-order signature consisting of a set of sorts $\Ss$ as well as a set of relation symbols $\Rr$ and function symbols (includes constant symbols) $\Ff$. Function and relation symbols are also equipped with a signature in the usual way: a function $f$ has a signature of the form $(\tau_1 \times \tau_2 \times \cdots \times \tau_n) \rightarrow \tau$ where $n$ is its arity and $\tau_i,\tau \in \Ss$, and a relation symbol $\rel$ has a signature of the form $\eta_1 \times \eta_2 \times \cdots \times \eta_m$ where $m$ is its arity and $\eta_i \in \Ss$.

Our annotations use a set of \emph{observer methods} corresponding to a class which are pure, side-effect-free, terminating functions on objects. These do not change the state of the class or of the object, and are used to measure properties of an object. 
For example, in a stack class, a method that returns the element
stored on the top of the stack (without popping it) would be an
observer method. Another example is a method that returns the size of
the stack. Observer methods can be used to write rich
annotations/specifications. For example, we could say that after
popping an element, the size of a stack object reduces by 1 (formally
\revision{$\mathit{size}'(o) = \mathit{size}(o) -1$}).
Note that we treat the receiver object as the first parameter for the method, and use primes to denote the value of the observer method computed in the post-state\footnote{The reason observer methods can be primed is because their computation depends on the heap. When the heap is modified, the observer methods become logically distinct from what they were before the modification. This is a common modeling choice in verification involving defined functions over heaps.}.

The annotations we consider are formulas in multi-sorted first-order logic over the above vocabulary. We require well-formed formulas to only use primed observer methods on primed objects, i.e., one cannot evaluate an observer method that computes over the heap in the post-state on an object from the pre-state. We will typically focus on using quantifier-free formulas.

Invariants for the client program are written not only using (as free variables) the variables  in scope, but also using observer methods applied to object variables in scope, and combinations using the FO logic over the primitive types.

The syntax for contracts is more involved. Rather than have preconditions and postconditions for methods, we write contracts as postconditions that can relate pre- and post-states (in particular, postconditions that hold upon certain preconditions holding can be expressed as $\textit{pre} \Rightarrow \textit{post}$). Contracts for a library method $m$ are logical formulas involving (a) the receiver object variable $o$, 
(b) the formal parameters $\overline{p}$ for $m$ (which could involve objects), 
and (d) a variable $\mathit{ret}$ for the returned value. Contracts can use both primed and unprimed versions of observer methods. 
For example, a contract for $\mathit{insert}(o,k)$ into a container object $o$, can be 
$(\mathit{size}'(o) = \mathit{size}(o) + 1) \land \mathit{contains}'(o,k)$ where $\mathit{size}$ and $\mathit{contains}$ are observer methods. In our problem definitions and evaluation, we fix particular precise logics for annotations. \revision{We model exceptions as special sentinel values to denote error states and evaluate formulas over an error-aware semantics. This is similar to modeling exceptions using \texttt{Option}-type return values.}

Observe here that invariants for the client can involve the client's local variables, but may also need to talk about the state of the object heap even though it is only accessible through the library methods. The observer methods mediate this interface and provide a vocabulary that the client can use to state invariants. Correspondingly the library methods must provide contracts in the same vocabulary. For a formal treatment of modular annotations in OOP, see 
\cite{AptOlderogHoareLogic}.

We denote formulas with free variables $\overline{p}$ and that use observer methods $\overline{obs}$ as $\varphi[\overline{\mathit{obs}}](\overline{p})$, and when the observer methods are clear from context, \revision{as simply} $\varphi(\overline{p})$.

\mypara{Object-Oriented Programs and Verification Conditions} We consider an object-oriented client program $P$ that utilizes a set of classes $\Cc$ and set of library methods $\Ll = \{L_1, \ldots, L_m\}$. 
We assume that $P$ has a set of methods including a main method $M$ (we call this part of the program the client $C$), apart from the library methods $\Ll$.
Our client programs can perform the usual computations associated with objects, including creating new objects using constructors,
and manipulate objects using method calls of the class, with conditionals, iterations, and function calls to internal auxiliary methods.  
We also assume that the library methods do not call the client code.

Importantly, our client programs contain \emph{assertions}. The verification problem we are interested in involves showing that these assertions are valid assuming certain contracts for the library methods. We capture this aspect of the problem by modeling the validity of the assertions with respect to contracts as the validity of a verification condition (VC) $\varphi$ in our logic described above. The VC $\varphi$ is stated over a set of free variables  $\overline{v}$ (modeling versioned variables for capturing basic blocks) as well as \emph{formula meta-variables} $\Ii= \{I_1,I_2,\ldots, I_n\}$ and $\Aa= \{A_1, A_2,\ldots, A_m\}$ corresponding to the invariants for the client code and the contracts for the library methods, respectively. We refer to client invariants in a broad sense, to include loop invariants and auxiliary method contracts. 
The signature and vocabulary of the invariants and the library contracts are as described above.

\revision{
We note that using contracts that refer to observer methods/pure functions is standard. Assertion frameworks established in the literature for Object-Oriented Languages like JML~\cite{JMLLogicBartJacobs2001,JMLTutorial} and Spec\#/Code Contracts~\cite{CodeContracts} allow contracts using observer methods. In OO languages, encapsulation naturally demands using observers to observe the internal state of objects.

When proving the client correct, we can model observer methods as fields of objects that are kept up to date across methods. 
Formally, we model them as uninterpreted functions that get updated according to contracts. Multiple objects do not pose any issue, and the modeling works in general. 

However, we acknowledge that in cases where clients and library methods make changes to a \emph{shared pointer-based heap} (not common in OO programming),
 our current contract language needs to be more expressive in order to be effective (e.g., include \emph{modifies sets} to facilitate frame reasoning, etc.) 
Although there is nothing that precludes our framework from handling such contracts, we do not utilize them in this work as we study applications that embrace encapsulation. We leave synthesizing such richer contracts clauses to future work.
}




We denote the VC with its parameters by $\varphi(\overline{v},\Ii,\Aa)$. The verification objective with respect to the client is to synthesize formulas for the predicates in $\Ii \cup \Aa$ (with the appropriate signature and in the logic) such that the VC is valid.\footnote{Verification conditions can be typically expressed as \emph{Constrained Horn Clauses}~\cite{seahorn,rusthorn}, and we will assume this form in our techniques. However, note that the observer methods in $\varphi$ are \emph{not} to be synthesized, which is not formally allowed in traditional formulations of CHCs. We show how we handle this when using CHC solvers later (see Section~\ref{sec:methodology:chcs})}.


\mypara{Test Generators} We use test generators to validate contracts
for the library methods with respect to the library implementation. We
model a test generator formally as a deterministic terminating oracle
$T(P, \revisiontwo{\mathit{mode}}, N, \widetilde{\Aa}; \testres)$ that
takes in a
program 
$P$ \revisiontwo{in the form of a client program $\clientprog$ and
  library $\lib$},
\revisiontwo{a mode (``modular'' or ``contextual'')}, a
\revisiontwo{library method $N$},
a set of contracts $\widetilde{\Aa} =
\{\widetilde{A}_1[\overline{obs},\overline{obs}'](\overline{x}_1),
\ldots,
\widetilde{A}_m[\overline{obs},\overline{obs}'](\overline{x}_m)\}$ for
the library methods with observer methods $\overline{obs}$ and free
variables $\overline{x}_i$ over which the contracts are expressed, and
a resource bound $\testres$. We use $\testres$ to model forced
termination since a testing strategy might go on forever. In practice,
since test generators can be nondeterministic, we determinize them by
fixing all random seeds.

\revisiontwo{In the modular mode,} the test generator does the following.
It 
creates objects and manipulates them using methods, with invented
parameters, to find valid objects, and generates test inputs for the
method $N$ and executes the method.
It looks for a \revisiontwo{violation of the contract for $N$},
by evaluating the terms in the contract, and substituting the values in
the formula. Specifically, it interprets the variable $\this$ to be
the receiver object, variables $\overline{p}$ to be the values
corresponding to the formal parameters, and $\mathit{ret}$ to be the
return value of the method. Importantly, it interprets the observer
method symbols in $\overline{obs}$ to be the functions defined by the
respective method definitions given in the code. It also executes
unprimed methods in the pre-state and their primed versions in the
post-state. 
It substitutes the computed values for all these terms in the contract
for $N$ and evaluates whether the contract holds. If any of them do
not hold, the test generator raises a contract violation. 

\revisiontwo{In the contextual mode, the test generator essentially
executes the client program $\clientprog$ with different inputs it generates. }
Like in the modular case it checks for violations of the method contracts
whenever a library method is invoked in the client, and raises
violations accordingly.

\revisiontwo{In both modes,} the test generator can either return
$\mathit{OK}$ to say that it was 
not able to find a contract violation, or return
a method name $N'$ along with a valuation of inputs to $N'$, its
return value, and
observer function outputs, that evidence the violation of the contract
for the method.

\subsection*{Problem Statement}

We are now ready to define the main problem of study for which we develop algorithms in Section~\ref{sec:methodology}.

\begin{definition}[Modular Contract Synthesis for Verification Modulo Tested Libraries]
\label{defn:synth-mc}
Fix classes $\Cc$, 
library $\Ll$, a program $P$ with both client code $C$ and library code, with verification condition $\varphi(\overline{v},\Ii,\Aa)$, test generator $T$, and resource bound $\testres$. Fix also  
the signatures of the client invariants $\Ii$ and method contracts $\Aa$, as described above.

Synthesize a set of formulas $\widetilde{\Ii}$ corresponding to \revision{client invariants} $\Ii$ over the appropriate variables, as well as a set of contracts $\widetilde{\Aa}$ 
corresponding to the library contracts $\Aa$ over the appropriate variables such that:
\begin{enumerate}
    \item the VC substituted with the definitions for the invariant and contract predicates is valid, i.e.,  $\varphi(\revision{\overline{v}},\widetilde{\Ii},\widetilde{\Aa})$ is valid, and
    \item the test generator $T(P,\revisiontwo{\modmode},N,\widetilde{\Aa}; \testres)$ returns $\mathit{OK}$ for each library method $N$.\qed
\end{enumerate}
\end{definition}

A set of invariants $\widetilde{\Ii}$ and library contracts $\widetilde{\Aa}$ is said to verify a client modulo tested (modular) library contracts (\vmtlmod) if they satisfy the two conditions in the above definition\footnote{Observe that the definition of the \vmtlmod synthesis problem is similar to SyGuS problems which utilize a syntactically defined space of hypotheses, along with logical constraints involving the functions to be synthesized. In our case we want to synthesize definitions for the $\Ii$ and $\Aa$ where the hypothesis space for each is the set of FOL formulas over the appropriate variables, and our constraints are twofold, involving a verifier and a tester. }.


\section{Contextual Contracts}
\label{sec:contextualcontracts}

We study a new notion of contracts that emerges in the setting of verification modulo tested library contracts, which we call \emph{contextual contracts}.
Contextual contracts are essentially contracts for library methods that are stated \emph{at each call-site in the client where library methods are called}, with potentially different contracts at each call site. The meaning of such contracts is that the summaries they denote hold when methods are called by the given client at the respective call sites. Note that these are not modular contracts in the classical sense in that they need not hold in arbitrary contexts. More precisely, at each call site of the client, when the client makes a call to the library, it implicitly ensures a more restrictive precondition on \emph{the states of the library} as well as the parameters passed to the method when the call is made, and the contextual contract captures more precise summaries for such calls.

We recall the illustrated example from Section~\ref{sec:illustratedeg}. When a client inserts only nonnegative integers into a container class, and later retrieves them and computes their sum, the client can assert that this sum is always nonnegative. Proving this using classical contracts is certainly possible, but the contracts will be more complex. If there is an observer method such as \textit{min}, as discussed in Section~\ref{sec:illustratedeg}, then there are valid modular contracts \emph{coupled with stronger client invariants} that can prove correctness of the client.

However, when observer methods such as \textit{min} are missing, modular contracts can get much more complex to state. In this example, we could define a logical predicate $\textit{min}$, but such a definition on a linked pointer-based container class would have to involve recursive definitions and more complex logics such as separation logic, which make both verification (of the client) and testing extremely hard, 
A contextual contract, on the other hand,  is delightfully simple---it simply says that in the context of the client, removal of any member from the set will be nonnegative, and helps us show that the sum computed is nonnegative. This contract for \textit{remove} is not true for all clients, of course. 

Note that modular contracts are contextual contracts as well, as contracts that hold in all contexts hold for the client's contexts.
However, contextual contracts need not be modular, as the above example shows.


\subsection*{Why are Contextual Contracts Simpler?}
\begin{wrapfigure}{r}{0.45\textwidth}
\includegraphics[scale=0.4]{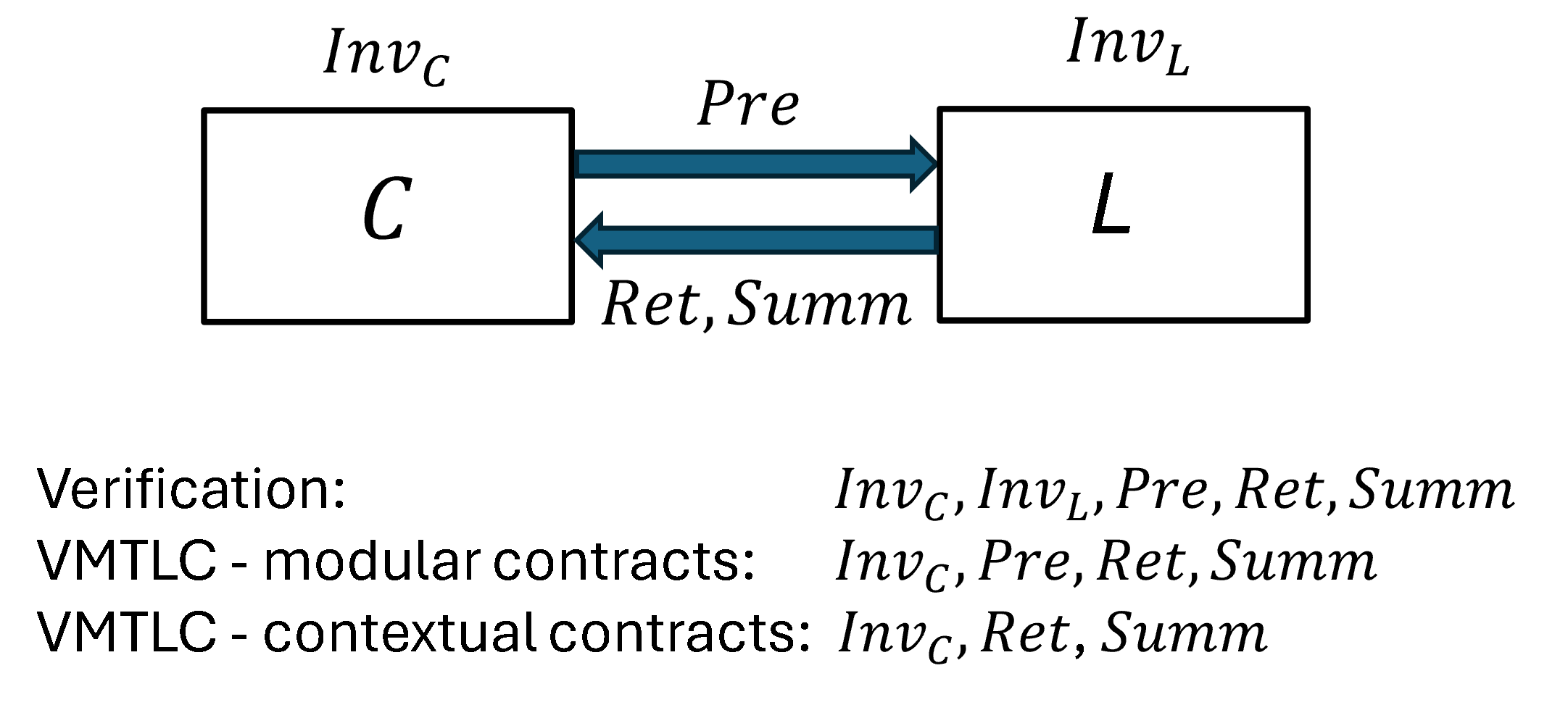}
\end{wrapfigure}
We now discuss precisely why contextual contracts are simpler than classical contracts.\\
Let $C$ be a client that using a library $L$, where $L$ exposes classes. Note that there is a \emph{global heap $H$ of objects} of the library that is modified by the library methods when they are called, and persist between calls. The client $C$ has references to instances but does not modify the heap directly.
Assume we want to prove a safety property of the client $C$.


~\\
The standard approach in formal verification is to establish:
\begin{itemize}
 \item An inductive invariant $Inv_C$ for the client that avoids the unsafe states of $C$. This invariant is expressed in terms of the variables of the client as well as  properties of the object heap $H$ maintained by the library that can be gleaned using observer methods.
 
 \item A precondition $Pre$ for when $C$ makes calls to $L$. This is a predicate on both the parameters passed to $L$ as well as the state of the object heap $H$ at the time of the call expressed using observer methods.
 
 \item A return condition $Ret$ that constrains the values returned by calls to methods of $L$
 
 \item $Summ$ that captures the 
 transformation of the object heap $H$ across method calls, expressed again using observer methods on the pre and post object heaps.
 
 \item An inductive invariant $Inv_L$ that is used to prove that $L$ does ensure $Ret$ and $Summ$ hold. This invariant can complex, involving the internal structure of the object heap, and may use aspects unobservable to the client.
\end{itemize}

In formal verification, we would prove (1) $C$ maintains the invariant $Inv_C$ and guarantees $Pre$ at calls to $L$, assuming $Ret$ and $Summ$ hold, and (2) the library $L$, assuming $Pre$, maintains the inductive invariant $Inv_L$, ensures return condition $Ret$, and ensures the summary $Summ$.

In verification modulo tested modular library contracts, we dispense with $Inv_L$, as we are not proving the library correct. We instead prove (1) above and \emph{test} that $L$, assuming $Pre$,  satisfies $Ret$ and summary $Summ$. Note that we need the annotations $Inv_C$, $Pre$, $Ret$ and $Summ$.

In verification modulo contextual contracts, we dispense with both $Inv_L$ and  $Pre$! 
We prove that $C$ maintains its invariant $Inv_C$ assuming $Ret$ and $Summ$ hold. And test that $L$'s contract in terms of $Ret$ and $Summ$ hold when tested in the context of the client $C$ (it need not hold \revision{when it is} called from arbitrary object heaps or arbitrary parameters). Note that we do not formalize $Pre$, as the precondition is implicitly the set of all parameters and object heaps at which $C$ calls $L$.

The advantage of using contextual contracts in the verification modulo testing setting is that it can be much simpler than modular contracts, as we do not have to precisely delineate the precondition $Pre$, and hence easier to synthesize. In fact, since $C$ is not proved to ensure $Pre$ formally, the verification task for the client is simpler, In particular, the return condition $Ret$, the summary $Summ$, and the inductive invariant $Inv_C$ can potentially be simpler as well.



\subsection*{Formal Problem of Verification Modulo Tested Libraries with Contextual Contracts}

We can now formally define the problem of verification modulo testing of libraries using contextual contracts:

\begin{definition}[Contextual Contract Synthesis for Verification Modulo Tested Libraries]
\label{defn:synth-cc}
Fix classes $\Cc$, library $\Ll$, a program $P$ with both client code
(with main method $M$) and library code, verification condition
$\varphi(\overline{v},\Ii,\Aa)$, test generator $T$, and resource
bound $\testres$. Fix also the client invariants $\Ii$, and method
contracts $\Aa$ (containing a contract for each call-site in client
code that makes a call to the library). \revisiontwo{We assume the
  VC $\varphi$ for the client is also in terms of these contextual
  method names.}

Synthesize a set of formulas $\widetilde{\Ii}$ corresponding to the client invariants $\Ii$ over the appropriate variables, as well as a set of formulas $\widetilde{\Aa}$ corresponding to the library contracts $\Aa$ over the appropriate variables, such that:
\begin{enumerate}
    \item the VC substituted with the invariants and contracts $\varphi(\overline{v},\widetilde{\Ii},\widetilde{\Aa})$ is valid, and
    \item the test generator $T(P,\revisiontwo{\conmode},M,\widetilde{\Aa}; \testres)$ returns $\mathit{OK}$.\qed
\end{enumerate}
\end{definition}

Note that in condition (2) above the method under test is the client itself (with entry point as main method $M$ of client), i.e., we require that the library contracts hold when calls are reached via the main method of the client. There are no individual ``unit'' testing requirements for each library method (in contrast to verification modulo testing of \emph{modular} library contracts; Definition~\ref{defn:synth-mc}). 
This is the sense in which the contracts are \emph{contextual}: they are only required to hold on inputs and library states that can occur in a context of the client program's calls to the library. \revision{Note that we are still limited to having one contract per call site (independent of the call stack).}

\revision{
\subsection{The Added Value of VMTLC with Contextual Contracts}
\label{subsec:VMTLCValue}
The argument as to why VMTLC with modular contracts adds value to just testing alone is that it must surely be hard to prove a wrong program correct by
generating spurious contracts for libraries that pass independent testing and provide inductive invariants that prove the client correct modulo these contracts. Of course, there are programs where this can happen, as testing of contracts is incomplete.

Contextual contracts help prove more correct programs correct (as described above) but they also can validate more incorrect programs. The reason is that there is no explicit precondition stated under which the library is tested independently, and the only tests of libraries occur in the context of testing the entire program, with the client. 

For example, let's assume that there is a rare event in the client that testing cannot reveal. An example code for a rare event is:\\
\centerline{ \texttt{ int x:=0; while (*) \{ int v:= nondet[1,10]; x :=x+v \}}
}
with the rare event being say ${\tt x}=1729$ after this code.
This rare event would be hard to find by fuzz testing (or other forms of testing); for the sake of argument in this section, let us assume this is so.

Now consider a client that manipulates a set--- it first adds only positive numbers to the set, except that it adds a negative number to the set when the rare event happens. The client then removes the elements from the set, and asserts that each element is positive. 

This program is incorrect, and testing will not find an error. However, a spurious contextual contract that says that removing elements from the set will always result in positive elements will pass the tester that tests the client and the library together. And using this spurious contract, we can prove the client correct. 

Note that in this case, the spurious contract will not pass as a modular contract, as testing a modular contract allows the tester to add negative numbers, which will easily show the contract to be incorrect.
A modular contract that has an explicit precondition saying elements added are positive will not work either as it would be impossible to prove that the client meets this precondition.

Given the above, it is natural to wonder if VMTLC with \emph{contextual contracts} adds any value to testing.
We do believe that spurious contextual contracts may  occur more often than spurious modular contracts, but posit that they do add value to just testing.

Here is an example.
Let {\tt foo} be a function that has access to a static/global variable $v$ (initially $0$). For an input integer $x$, it returns $2$ when $x=v$ and increments $v$ by $3$, and otherwise, returns $0$.
Consider the program:\\
\centerline{\tt 
x:= 0; while (*) \{  y:=nondet();  z:= foo(y);
  x:= x + z;
\}; assert (x != 1220);
}

The above program is incorrect, but it would be hard for testing to find an error. However, note that any contextual contract (or modular contract) for {\tt foo} that allows the return value to be $2$ cannot help prove the program correct. Furthermore, testing will easily reject any contract that doesn't allow the return value to be $2$, even when tested as the entire program (one of the nondeterministic values that {\tt foo} is called with needs to be $0$, in some test case). Consequently, VMTLC with contextual contracts (or modular contracts) will not prove this program correct. Also, note that if {\tt foo} instead always returned values that are a multiple of $3$, contextual (and modular) contracts exist that prove the program correct (with a loop invariant that $x$ is a multiple of $3$).

We hence believe that contextual contracts may also add value to testing, as it continues to be the case that finding spurious contracts that enable inductive invariants of the client to prove it correct will be hard to find in practice. 
}

\section{A Solution to the VMTLC Problem}
\label{sec:methodology}
In this section, we describe our solution to the VMTLC problem. We begin with an overview of our algorithm, followed by a discussion of our design choices for input verification conditions and the helper methods used in our algorithm, and finally present the algorithm.

\subsection{Overview}
\begin{figure}
  \begin{center}
    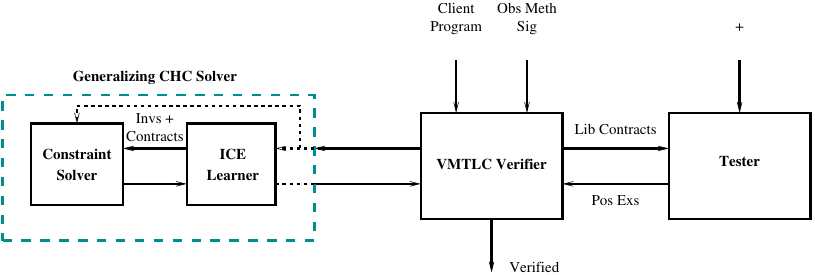
\caption{Schematic diagram of our proposed solution.}
\label{fig:approach-schematic-ICE}
  \end{center}
\end{figure}

Fig.~\ref{fig:approach-schematic-ICE} shows the schematic diagram of
our general proposed technique. It takes as input a client program
that calls library methods ($C$), as well as verification conditions
generated for the client with annotation and contract
holes ($CHC(C)$). Additionally, the observer methods of the library are also
passed as input ($O$). These observer methods are used in the contracts of
the library method.

Our technique then synthesizes contracts for the library methods and
invariants for the client program in a counterexample guided loop. In
each iteration of the loop, contracts and invariants are first
synthesized by a \emph{generalizing CHC solver} (details to follow in
Section~\ref{sec:methodology:gchc}) such that the verification
conditions are valid. This is followed by testing of contracts: either
the library methods for modular contracts, or the client program for
contextual contracts. If the testing fails, the test cases that cause
the failure are passed as positive examples to the generalizing CHC
solver. In the next iteration, the generalizing CHC solver learns
contracts for methods (ensuring that the positive examples are
included), along with invariants. This loop continues until the solver
synthesizes contracts and invariants that verify the client, and where
the contracts also pass the testing engine.

\subsection{Constrained Horn Clauses as Verification Conditions}
\label{sec:methodology:chcs}

We assume a \emph{verification condition generator} that works on the client program to produce verification conditions in the form of  constrained Horn clauses (CHCs). Our technique is agnostic to the semantics of the programming language, the  specification logics, frame reasoning, etc., as VC generation into CHC constraints takes these into account, and our framework takes CHCs as input. CHCs have  become popular as a symbolic intermediate representations of correctness of programs, and verification conditions for CHCs have been developed for several programming languages and program reasoning methods have been developed for CHCs~\cite{arrfreq,rusthorn, seahorn,KahsaiRSS16,lpar/KahsaiKRS17,Eldarica,pldi/GrebenshchikovLPR12,pldi/ZhuMJ18,multifreq,prabhu2018efficiently}. A CHC is a first-order logic formula over a fixed set of uninterpreted relations, and a system of CHCs is a finite conjunctions of such formulas. A solution to a CHC system maps uninterpreted relations to predicates from a fixed logic. When a CHC system represents a program's verification conditions, a solution to the system is a proof of the program's correctness.

Formally, for a fixed set of sorts $\Ss$, a set of relation symbols $\Rr$, and function symbols $\Ff$, a CHC can be of one of the following forms:
\begin{align*}
 \forall \vx_0 \such \big(\psi(\vx_0) & \implies  \rel_0(\vx_0)\big) \tag{fact}\label{eqn:fact} \\  
 \forall \vx_0,\ldots,\vx_{k+1} \such \big( \bigwedge\limits_{0\le i \le k}\!\!\rel_i(\vx_i) \, \land\, \psi(\vx_0, \ldots ,\vx_{k+1}) & \implies  \rel_{k+1}(\vx_{k+1})\big)  \tag{inductive}\label{eqn:ind}  \\  
 \forall \vx_0,\ldots,\vx_{k} \such \big(\bigwedge\limits_{0\le i \le k}\!\!\rel_i(\vx_i) \, \land \, \psi(\vx_0, \ldots ,\vx_{k}) & \implies  \false \big) \tag{query}\label{eqn:qu}
\end{align*}
where, for each $i$, $\rel_i$ is a relation in $\Rr$ 
of type $\eta_1 \times \eta_2 \times
\cdots \times \eta_{m_{\rel_i}}$, where $m_{\rel_i}$ is the arity of
$\rel_i$ and each $\eta_i \in \Ss$. Further, $\vx_i$ represents the vector
of variables $(x_1, \ldots, x_{m_{\rel_i}})$, with each $x_i$ of sort
$\eta_i$. Also, $\psi$, called a \emph{constraint}, 
is a formula in conjunctive normal form without uninterpreted
relations.

For a given client program, $\Ss$ consists of primitive sorts (like
boolean and integer) and designated sorts of classes used in the
program (like \texttt{Set} or \texttt{BinarySearchTree}), $\Rr =
\{I_1,I_2,\ldots, I_n\} \cup \{A_1, A_2,\ldots, A_m\}$, containing
relations representing invariants and library method contracts,
respectively, and $\Ff$ consists of operations and constants used in
the program. The CHCs then encode the proof obligations of a
Floyd-Hoare style proof, 
with constraints in them corresponding to program statements. In CHCs, the
relations $I_1,I_2,\ldots, I_n$ will be over variables that are in
scope, and the relations $A_1, A_2,\ldots, A_m$ will be over the called
object, and arguments and return values of the corresponding methods
they are representing.

\begin{example}
\begin{figure}
\begin{small}
\begin{align*}
\consetinit(\varset) \land i=0 &\implies \invoneset'(i,N,\varset) &(\varphi_{C_1}) \\ 
\invoneset(i,N,\varset) \land i < N \land \mathit{v1} \ge 0 \ \land \consetinsert(\varset, \mathit{v1}) \land  i' = i + 1 &\implies \invoneset'(i',N,\varset) & (\varphi_{C_2}) \\ 
\invoneset(i,N,\varset) \land \neg (i < N)  \land \mathit{sum} =0 &\implies
\invtwoset(N, \varset, \mathit{sum})  & (\varphi_{C_3})\\ 
\invtwoset(N,\varset,\mathit{sum}) \land \neg \consetempty(\varset) \land \consetremove(\varset, \mathit{v2}) \land \mathit{sum}' = \revisiontwo{\mathit{sum} + \mathit{v2}} &\implies \invtwoset'(N,\varset, \mathit{sum}') & (\varphi_{C_4}) \\ 
\invtwoset(N, \varset, \mathit{sum}) \land \consetempty(\varset) \land \mathit{sum} < 0 &\implies \false &(\varphi_{C_5})
\end{align*}
\end{small}
\caption{CHCs for the client program from Fig.~\ref{fig:set-client}.}
\label{fig:bst-chc}
\end{figure}

    Consider the CHCs in Fig.~\ref{fig:bst-chc}, corresponding to a
    Floyd-Hoare proof of the client program of
    Fig.~\ref{fig:set-client}(a). Here, $\Rr =
    \{\invoneset, \invtwoset\} \cup \{\consetinit, \consetinsert,
    \consetempty, \consetremove\}$, where $\invoneset$ and
    $\invtwoset$ are the loop invariants\footnote{$\invoneset'$ and
    $\invtwoset'$ are the same relations with slightly different
    interpretations; details to follow soon.}, $\consetinit$
    represents the constructor, and $\consetinsert$, $\consetempty$ and
    $\consetremove$ represent the class methods $\fnsetinsert$,
    $\fnsetempty$ and $\fnsetremove$, resp. Since variables of types
    \texttt{int} (variables $N, sum, v1, v2$), and \texttt{Set}
    (variable $\varset$) are used in the program, $\Ss =
    \{\mathbb{Z},\mathbf{Set}\}$. Further, the signature of
    $\consetinit$ will be $\mathbf{Set}$, as it initializes the object
    $\varset$, $\consetinsert$ will be $\mathbf{Set} \times
    \mathbb{Z}$, as it is called on the $\varset$ object and takes an
    \texttt{int} variable as input. Similarly, the signature of
    $\consetempty$ will be $\mathbf{Set}$, and $\consetremove$ will be
    $\mathbf{Set} \times \mathbb{Z}$. The invariant relation
    $\invoneset$ will have the signature $\mathbb{Z} \times \mathbb{Z}
    \times \mathbf{Set}$, as the variables $i, N, \varset$ are in
    scope of the loop, while $\invtwoset$ will have the signature
    $\mathbb{Z} \times \mathbf{Set} \times \mathbb{Z}$, corresponding
    the variables $N, \varset, sum$. These relations are used in the
    CHCs $\varphi_{C_1}$ to $\varphi_{C_5}$, where CHCs
    $\varphi_{C_1}$ to $\varphi_{C_4}$ are of type ~\ref{eqn:ind} and
    correspond to the initial and inductive conditions of the loops,
    while $\varphi_{C_5}$ is of type ~\ref{eqn:qu}.
    \end{example}

For a system of CHCs over $\Rr$, a map of interpretations $\MM$
assigns to each relation symbol $\rel \in \Rr$ an interpretation
$\theta(x_1,\ldots,x_{m_{\rel}})$, where $\theta$ is a formula in a
fixed logic. We use a quantifier-free fragment of a
multi-sorted first-order logic of annotations (defined in
Section~\ref{sec:prelim}). This logic will have a finite set of
observer methods and their primed versions as uninterpreted functions
of fixed sorts.

For a CHC $\varphi_C$ and a map $\MM$, we can obtain a formula
$\varphi_C(\MM)$ by replacing all the relations
$\rel_i(y_1, \ldots, y_{m_{\rel_i}})$ in $\varphi_C$ by
$\MM[\rel_i](y_1/x_1, \ldots,
y_{m_{\rel_i}}/x_{m_{\rel_i}})$\footnote{For primed relations like
$\invbst'$, the interpretation $\MM[\invbst]$ is used, except that
uninterpreted functions corresponding to observer methods in
$\MM[\invbst]$ are replaced by their primed versions.}. If this
formula is valid, then $\MM$ is a \emph{solution} to $\varphi_C$ and
$\varphi_C$ is said to be \emph{satisfiable}.
$\MM$ is a \emph{solution} to a
system of CHCs if it is a solution to all the CHCs in the system. If
it is not a solution to a CHC $\varphi_C$, a counterexample to
validity (a ground formula in the logic of $\MM$) can be
obtained. Such examples are classified as \emph{positive}
if $\varphi_C$ is of type ~\ref{eqn:fact}, \emph{Horn implication}
if $\varphi_C$ is of type ~\ref{eqn:ind}, and \emph{negative}
if $\varphi_C$ is of type ~\ref{eqn:qu}.

\subsection{Need for Generalizing CHC Solvers}
\label{sec:methodology:gchc}

Recall that in our setting the input $\varphi$ is a system of CHCs
over $\Rr = \{I_1,I_2,\ldots, I_n\} \cup \{A_1, A_2,\ldots, A_m\}$,
which are relations representing the client program's invariants and
library method contracts, respectively. The positive examples returned
by the tester in each round in our framework can also be encoded as
CHC constraints (these constraints will be \emph{fact}
constraints where the relation on the right side of the implication
is a contract relation $A_i$).

Our technique calls for  a \emph{generalizing CHC solver} to synthesize a solution to $\varphi$, as opposed to an arbitrary CHC solver. 
Note that the contracts $A_i$ are only constrained from \emph{below}
by the positive counterexamples returned by the tester.
An arbitrary CHC solver could choose to solve the CHC by overfitting
this set of positive examples (say by insisting that the $A_i$'s cover
precisely the positive counterexamples). This would cause repeated
calls to the tester and cause divergence, where we never
find the right contracts.
Hence we require that the CHC solver generalizes the library method
contracts $A_1, \ldots, A_m$. One common way to generalize is to bias
the search towards \emph{small} logical expressions for each $A_i$ in
the annotation logic, which automatically forces generalization. 

\subsection{ICE Learners as Generalizing CHC Solvers}
\label{sec:methodology:ice}

We use the data-driven learning framework called Implication
Counterexample (ICE)
learning~\cite{ice,icedt,DBLP:journals/pacmpl/EzudheenND0M18,tacas/ChampionC0S18}
as the basis of a generalizing CHC solver. Learners in this framework find a
general solution by learning from positive, Horn implication, and
negative examples, which are counterexamples to the satisfaction of
CHCs. At a high level, ICE learners operate in an incremental
loop. In each iteration, a candidate solution is learned using
examples from previous iterations. This candidate solution is then
checked to see whether it satisfies all the CHCs. If not,
counterexamples to satisfaction are collected and used in
subsequent iterations to refine the candidate solution.

\alghice

Algorithm~\ref{alg:hice} provides a formal description of a
generalizing CHC solver built on the ICE
learner framework. It takes as
input the CHCs $\varphi$ and positive examples $\Pex$ from the test
generator. The algorithm first initializes the set of
examples ($\Ex$) with the input set of positive examples
(line~\ref{alg:hice:init}).
Subsequently, in each
iteration of the loop, the algorithm checks whether there exists a CHC
$\varphi_C$ for which the current candidate $\MM_{cand}$ is not a
solution (line~\ref{alg:hice:ex}). If such a $\varphi_C$ is found, a
counterexample is generated (line~\ref{alg:hice:cex}). Recall that the
counterexample is a ground formula in the logic of the solution. It
will be classified as a positive example if $\varphi_C$ is of type
~\ref{eqn:fact}, a Horn implication example if $\varphi_C$ is
~\ref{eqn:ind}, and a negative example if ~\ref{eqn:qu}. A new
candidate solution is then learned that is not only consistent
with the examples, but also generalizes the solution
(line~\ref{alg:hice:dt}). This loop repeats till $\MM_{cand}$ becomes
a solution to the system $\varphi$.

\medskip
We now present different realizations of the learning algorithm that
synthesize generalized candidate solutions which are consistent with a given
set of examples.

\subsubsection{\hice}
An ICE learner, \hice~\cite{DBLP:journals/pacmpl/EzudheenND0M18} uses
a decision tree learning algorithm to synthesize solutions, extending
standard decision tree algorithms for positive/negative samples to
work on ICE samples. A candidate solution is synthesized by learning a
decision tree for each relation, and works in polynomial time. The
decision tree is guaranteed to be consistent with the ICE samples. The
decision trees constructed in this way have base formulas from a fixed
template (e.g.\@ octagonal constraints over variables). As a result,
decision trees correspond to boolean combinations of these base
formulas. Also, the construction favors small decision trees by using
heuristics such as entropy, which often leads to generalized formulas. 

\subsubsection{Large Language Models (LLMs)}
Transformer-based neural networks realizing Large Language Model (LLM) have recently been found to be useful in program verification tasks, such as invariant and precondition synthesis~\cite{kamath2023finding,cao2025clause2inv,wen2024enchanting,king2025llm}. Furthermore, they have been trained on coding tasks~\cite{team2024gemini,achiam2023gpt}, many of which use data-structure libraries. Motivated by these, we use an LLM as an ICE learner.

Our algorithm interacts with the LLM through textual prompts (the
prompt templates can be found in
Appendix~\ref{sec:app:llmprompts}). The initial prompt includes the
input client program ($C$), its CHC encoding ($\varphi_C$), and a
textual description of the task (see~\ref{sec:app:prompt:init}). The
algorithm expects the LLM to return a
candidate solution to the formula variables in $\varphi$, which
includes invariants and library method contracts. If the candidate is
not a solution or subsequent testing fails, a repair prompt is
generated, which includes the examples, and sent to the LLM, expecting
it to repair the solution (prompt templates in~\ref{sec:app:prompt:repair1} and  ~\ref{sec:app:prompt:repair2}).

\subsubsection{Combination of \hice and Large Language Models}
 \revision{In this
approach, initially the LLM is queried to get atomic formulas (prompt template in ~\ref{sec:app:prompt:hllm}). These are passed to the \hice learner as a template. Subsequently, \hice is used as the ICE learner.} It now considers the normalized atomic formulas as node attributes, in place of the fixed templates used previously, and learns decision trees.

\subsection{Test Generator}
In our algorithm, the test generator takes as input library contracts
present in the solution synthesized  by the learner and tests them
against the library implementation. This step is necessary as a
solution of the CHC system can overly restrict the contracts of
library methods. Our algorithm expects the test generator to check
such solutions by generating test cases and returning the failing test
cases as positive examples. In
turn, the learner generalizes the contracts by using these positive
examples. For example, consider the solution to the CHCs in
Fig.~\ref{fig:bst-chc}, which maps all the relations to $\false$ (which
vacuously satisfies all the CHCs). In this case, the test generator
should produce positive examples for library methods which violate
the specification $\false$ as a method contract.

\algtest

A formal description of the test generator is presented in
Algorithm~\ref{alg:test}.
\revisiontwo{
The algorithm takes as input the client program $\clientprog$, the
library $L$, the mode of verification (modular/contextual), the
library method $N$ to test (for use in modular mode), contracts for the
library methods, and a testing resource bound.
The call to \textsc{ConstructProgram} constructs a program (\testprog)
which will read its inputs from an input buffer $\inputbuf$ and test
the appropriate method contracts.
In the modular mode, $\testprog$ creates library objects necessary to
invoke the method $N$, sets up the state of the these objects by
calling library methods on them (the operations to call and
the arguments to pass are determined by the input buffer
$\inputbuf$). It then invokes method $N$ on the appropriate object
(once again with arguments determined by $\inputbuf$). Before and
after this invocation, it records the values of the observer methods
on the pre/post state of the object, along with the arguments and
return value of $N$, in $\recval$.
Finally, it asserts the contract for $N$ (i.e.\@ $\MM(N)$) with the values
from $\recval$ plugged in.
If the assertion fails it returns $\Tok=\false$, along with the method
name $N$ and $\recval$. If the assertion passes it returns $\true$.
}
\revisiontwo{If the mode is contextual, $\testprog$ is essentially
the client program $\clientprog$ with instrumentation for each
contextual method call, similar to the modular mode.
}
\revisiontwo{Then, in a loop, the test generator
algorithm initializes $\inputbuf$
(line~\ref{alg:test:init}), and executes $\testprog$ on it.
If $\testprog$ returns an assertion failure, the corresponding method
name $N'$ and valuation $\recval$ are returned.}
The loop continues until $\testres$ is
exhausted. If no assertion violation is found in the loop, the
algorithm returns success (line~\ref{alg:test:retsuc}).

\subsection{Complete Algorithm}

\algmain

Algorithm~\ref{alg:main} gives a formal description of our approach to
find contracts in the verification modulo testing setting.
\revisiontwo{It takes as
input a client program $\clientprog$, along with 
the library $L$, a string $\mathit{mode}$ representing the
verification mode (contextual/modular), corresponding CHCs of the program
($\varphi$), and resource constraints on the test generator
($\testres$).
}
The algorithm operates in a counterexample guided synthesis loop until
it finds a solution to $\varphi$ (variable $\CHCok$ is $\true$) that
also passes testing (variable $\Tok$ is $\true$). In the first
iteration of the loop, a generalizing CHC solver (helper method
~\textsc{ICE-CHCSolver}) synthesizes a map of interpretations ($\MM$,
line~\ref{alg:main:gcs}) that is a solution to the CHCs
$\varphi$. This is followed by testing, where the library contracts
in $\MM$ are tested. 
\revisiontwo{
If the mode is contextual, the contracts are tested through the client
program $\clientprog$ (line~\ref{alg:main:tc}).
If the mode is modular, the contracts for each library method $N$ are
checked separately (line~\ref{alg:main:tm}) via a test harness.
Whenever testing fails 
the corresponding valuations ($P$) are
added to the set of positive examples $\Pex$
(lines~\ref{alg:main:pc} and \ref{alg:main:pm}).
}
In the subsequent iterations, \textsc{ICE-CHCSolver} will make sure that the
solution it synthesizes to $\varphi$ also \emph{generalizes} positive
examples in $\Pex$. Algorithm~\ref{alg:main} always terminates with a
solution if the following conditions are satisfied: 1) the logic of
the CHC solutions is finite, 2) a general solution exists within the
logic, and 3) \textsc{ICE-CHCSolver} correctly classifies examples.

\section{Implementation} 
We implement our framework (Algorithm~\ref{alg:main}) in a tool named $\vmtlc$, in Python. $\vmtlc$ orchestrates the communication between the learning algorithms and the constraint solvers in the framework. In particular, our implementation offers the following ICE learners: (1) \hice~\cite{DBLP:journals/pacmpl/EzudheenND0M18}, a decision tree learner, (2) an LLM-based ICE learner built using Gemini v2.5 Pro and appropriate prompts, (c) and a combined \hllm learner that uses the decision tree learner for synthesizing client invariants \revision {and library contracts using atomic formulas generated by Gemini v2.5 Pro.} We use \zt~\cite{Z3} to check verification conditions for the client. The test generator is implemented using the \afl\footnote{\url{https://github.com/AFLplusplus/AFLplusplus}} fuzzing framework.

The \hice learner synthesizes expressions in a grammar that is a boolean combination of octogonal constraints over integer variables and booleans. The variables are input parameters and values of observer functions. In the case of the combined learner, the LLM is asked for base expressions which are incorporated into the grammar.

To use \afl for testing, we create a test harness that (a) creates objects and manipulates them using methods from the class as dictated by an input controlled by \afl, and (b) calls the method under test instrumented with the candidate contracts as runtime assertions. When a contract violation is encountered, we get a counterexample. Note that the testing setup is different for modular and contextual contracts; modular contracts are tested directly with the library methods as the methods under test, and contextual contracts are tested via the client. 




We provide different prompts to the LLM for synthesizing modular and contextual contracts (see Appendix~\ref{sec:app:llmprompts}). The prompt contains (a) the client code, (b) signatures of library methods, (c) the CHCs corresponding to the VC, (d) ICE counterexamples from the verifier and positive counterexamples from the test engine, and (e) instructions for the LLM for the contract and invariant synthesis problems, including the expected output format.




\section{Evaluation}
\label{sec:eval}

In this section we describe our benchmark suite and the research questions we study. 


\subsection*{Benchmarks}
Our benchmark suite consists of 43 C++ client programs that utilize various popular open-source data structure libraries. These client programs are primarily adapted from unit tests into more challenging clients for verification. 
We derive 19 benchmarks from Folly\footnote{\url{https://github.com/facebook/folly}} (libraries used in Facebook), 9 from Abseil\footnote{\url{https://github.com/abseil/abseil-cpp/}} (libraries used in Google), and 1 from the Android source-code\footnote{\url{https://cs.android.com/android/kernel/superproject}}. Additionally, we incorporated 14 benchmarks from SV-COMP Java category ~\cite{beyer2025improvements}, which already had client and library implementations. We rewrite these benchmarks into C++. Overall, these benchmarks include libraries with lines of code ranging from 100 to 3000, client programs with 2-6 library methods. 6 of our benchmarks are $\sim$3000 LoC. A detailed description of the suite can be found in Table~\ref{tab:benchmarks} in Appendix~\ref{sec:app:benchmark}. 

Our benchmark suite is beyond the scope of the state-of-the-art in full automated verification. More precisely, we tried to verify our benchmarks using SeaHorn~\cite{seahorn}, and all but 7 of them failed to verify or returned spurious counterexamples. The 7 benchmarks contained libraries on the smaller side in our suite (< 200 LoC). In fact, even when given contracts for the library methods that would prove the client correct, SeaHorn failed to verify.

\revision{Our suite contains benchmarks that are challenging both in size and in the complexity of the verification problem. For example, the FlatHashMap4 benchmark, has a client with $\sim$30~LoC and a library with $\sim$3000~LoC. The library is adapted from similar classes found in both the Abseil and Folly repositories, and the client is adapted from a test harness that captures a common usage pattern of the data structure. 
The ProcessQueue benchmark is another interesting example, adapted from SV-COMP and derived from a real-world scheduling application. The client 
($\sim$20~LoC) utilizes a library ($\sim$100~LoC) that implements a waiting process queue, tracking time-to-wake for processes, returning those that are due.
SeaHorn was not able to verify either of these benchmarks automatically given more than an hour.


}

\subsection*{Research Questions }
We study the following research questions:
\begin{table}[t]
  \centering
  \scriptsize
  \setlength{\tabcolsep}{3pt}
  \renewcommand{\arraystretch}{1.1}
\begin{tabular}{l|rr|rr|rr|rr|rr|rr}
    \multirow{3}{*}{Benchmark} &
      \multicolumn{4}{c|}{\hice} &
      \multicolumn{4}{c|}{\hllm} &
      \multicolumn{4}{c}{\llm} \\ \cline{2-13}
      & \multicolumn{2}{r|}{Modular} & \multicolumn{2}{r|}{Contextual} &
        \multicolumn{2}{r|}{Modular} & \multicolumn{2}{r|}{Contextual} &
        \multicolumn{2}{r|}{Modular} & \multicolumn{2}{r}{Contextual} \\ \cline{2-13}
                         & E+I       & T(m)      & E+I     & T(m)     & E+I     & T(m)       & E+I      & T(m)    & E+I      & T(m)  & E+I     & T(m)        \\ \hline
    AlternatingList      & 8+311     & 12.00     & 5+154   & 3.86     & 6+95    & 2.67       & 1+70     & 2.27    & 2+3      & 3.12  & 1+1     & 0.92        \\ \hline
    AtomicHashMap1       & 2+3       & TO        & 4+9     & TO       & 5+199   & TO         & 4+5      & TO      & 5+7      & 10.68 & 2+4     & 11.38       \\
    AtomicHashMap2       & 3+31      & TO        & 5+212   & TO       & 4+87    & TO         & 6+117    & TO      & 4+5      & 8.95  & 2+3     & 6.28           \\
    AtomicHashMap3       & 2+16      & TO        & 2+15    & TO       & 2+7     & TO         & 2+8      & TO      & 8+10     & 20.77 & 1+1     & 1.83        \\
    AtomicHashMap4       & 5+140     & TO        & 3+16    & 1.35     & 5+65    & TO         & 3+50     & 1.29    & 6+7      & 10.72 & 2+2     & 4.33            \\
    AtomicHashMap5       & 4+58      & TO        & 5+51    & 3.36     & 6+98    & TO         & 17+124   & TO      & 2+2      & 8.92  & 2+2     & 4.25           \\ \hline
    AtomicLinkedList1    & 5+225     & TO        & 4+134   & 2.49     & 4+125   & TO         & 4+38     & 2.17    & 2+3      & 2.38  & 1+2     & 1.55        \\
    AtomicLinkedList2    & 2+12      & TO        & 2+12    & TO       & 2+7     & TO         & 3+183    & TO      & 15+25    & TO    & 15+21     & TO         \\ \hline
    BinaryHeap1          & 2+14      & TO        & 2+14    & TO       & 2+8     & TO         & 2+8      & TO      & 15+31      & TO    & 7+9     & 16.97        \\
    BinaryHeap2          & 3+47      & 4.31      & 4+14    & 2.23     & 3+43    & 4.96       & 4+13     & 2.20    & 1+1      & 1.58  & 2+2     & 3.18        \\ \hline
    BinaryTree           & 5+88      & 6.90      & 4+48    & 2.21     & 3+7     & 6.72       & 3+28     & 1.30    & 2+1      & 3.12  & 3+5     & 2.25        \\
    BlueWhiteList        & 5+42      & 3.36      & 4+27    & 2.20     & 6+57    & 4.69       & 4+12     & 2.14    & 1+1      & 0.88  & 1+1     & 1.15        \\
    Calendar             & 5+199     & 5.58      & 6+139   & 4.96     & 5+32    & 3.60       & 1+20     & 3.23    & 1+1      & 1.3   & 1+1     & 1.80        \\
    DLL\_Circular        & 9+144     & 10.26     & 6+74    & 4.63     & 9+52    & 6.67       & 6+25     & 4.33    & 1+1      & 0.96  & 1+1     & 1.07        \\
    DLL\_Token           & 5+335     & TO     & 3+157   & 5.08     & 2+18    & 14.90      & 6+150    & 5.70    & 1+1      & 1.37  & 3+3     & 2.60        \\ \hline
    FlatHashMap1         & 7+182     & 13.52     & 5+131   & 3.49     & 10+58   & TO         & 4+28     & 2.18    & 4+4    & 6.22    & 1+1     & 1.07       \\
    FlatHashMap2         & 11+273    & TO        & 2+6     & 3.43     & 9+54    & 16.33      & 5+61     & 3.28    & 9+11    & 16.37    & 1+1     & 0.83       \\
    FlatHashMap3         & 8+290     & TO        & 4+144   & 3.67     & 5+142   & TO         & 4+35     & 2.24    & 15+19    & TO    & 1+2     & 2.93     \\
    FlatHashMap4         & 3+59      & TO        & 3+55    & 4.3      & 6+93    & TO         & 3+53     & TO      & 15+22    & TO    & 2+3     & 5.72     \\ \hline
    FlatHashSet          & 2+5       & TO        & 5+23    & TO       & 2+5     & TO         & 5+14     & TO      & 13+20    & 43.75    & 3+5    & 4.91      \\
    LruCache1            & 2+3       & TO        & 3+6     & TO       & 4+288   & 12.8       & 3+4      & TO      & 15+18    & TO    & 15+20   & TO           \\ \hline
    List (Max)           & 5+75      & 3.43      & 5+62    & 3.24     & 11+275  & 3.09       & 5+43     & 4.43    & 2+3      & 1.08  & 1+1     & 0.73        \\ 
    List (Min)           & 5+76      & 3.38      & 2+20    & 0.68     & 11+738  & 5.35       & 2+19     & 0.69    & 1+1      & 0.63  & 1+1     & 0.63        \\ \hline
    Multimap1            & 5+283     & TO        & 14+40   & 2.79     & 5+33    & 4.13       & 3+33     & 1.32    & 2+2      & 1.45  & 1+1     & 0.70        \\ 
    Multimap2            & 3+19      & TO        & 3+6     & TO       & 12+75   & TO         & 3+4      & TO      & 1+1      & 0.85  & 1+1     & 0.85        \\ \hline
    Multiset1            & 3+36      & TO        & 2+8     & TO       & 4+196   & 5.14       & 4+128    & 3.63    & 1+1      & 0.72  & 1+1     & 0.7         \\ 
    Multiset2            & 3+12      & TO        & 3+6     & TO       & 15+138  & TO         & 4+22     & TO      & 1+1      & 0.85  & 1+1     & 1.05        \\ \hline
    NormalFilterQueue    & 3+67      & 2.39      & 4+131   & 2.64     & 3+23    & 2.30       & 4+26     & 2.17    & 2+2      & 2.01  & 1+1     & 1.32        \\ 
    PriorityFilterQueue  & 3+67      & 2.48      & 3+95    & 1.59     & 4+45    & 4.46       & 7+61     & 2.21    & 2+2      & 2.45  & 1+1   & 2.24           \\ 
    ProcessQueue         & 3+24      & TO     & 5+211   & 3.00     & 4+127   & 4.88       & 5+95     & 2.62    & 13+23    & 40.97 & 1,2     & 1.87        \\ 
    RedBlackTree         & 2+3       & TO        & 4+20    & 1.56     & 11+57   & TO         & 4+18     & 1.55    & 2+2      & 5.70  & 1+1     & 1.83           \\ \hline
    SkipList1            & 2+3       & TO        & 4+228   & 9.68     & 2+18    & TO         & 5+244    & 3.22    & 2+3      & 7.07  & 3+3     & 4.38        \\ 
    SkipList2            & 3+17      & 2.71      & 4+16    & 2.17     & 3+4     & 2.76       & 4+8      & 2.13    & 2+3      & 4.02  & 3+4     & 4.82        \\ 
    SkipList3            & 3+11      & 2.72      & 4+14    & 2.13     & 3+11    & 2.71       & 4+8      & 2.13    & 2+2      & 4.23  & 2+2     & 3.50        \\ 
    SkipList4            & 2+15      & TO        & 3+84    & TO       & 2+7     & TO         & 2+7      & TO      & 2+4      & 6.87  & 1+3     & 3.54        \\
    SkipList5            & 3+21      & 2.74      & 4+34    & 2.17     & 3+20    & 2.30       & 4+16     & 2.16    & 3+3      & 6.72  & 2+2     & 3.77        \\ 
    SkipList6            & 3+32      & TO        & 4+198   & 7.20     & 3+5     & TO         & 4+60     & 2.23    & 3+5      & 8.23  & 2+4     & 4.50        \\ 
    SkipList7            & 4+52      & 4.70      & 4+26    & 2.15     & 3+19    & 3.40       & 4+13     & 2.13    & 2+5      & 3.67  & 3+5     & 7.62        \\ \hline
    Stack                & 6+330     & 11.61     & 6+254   & 5.34     & 5+163   & 18.07      & 6+71     & 4.40    & 2+3      & 2.70  & 1+1     & 1.63        \\ 
    StockOrder           & 3+72      & 1.54      & 5+90    & 3.36     & 3+38    & 1.51       & 4+49     & 2.13    & 2+6      & 5.93  & 1+1     & 0.95        \\ \hline
    TokenBucket1         & 3+186     & 5.00      & 6+229   & 4.20     & 3+30    & 4.34       & 1+14     & 1.16    & 2+3      & 2.55  & 1+1     & 1.03        \\ 
    TokenBucket2         & 3+84      & 2.90      & 4+97    & 1.94     & 3+50    & 2.15       & 7+45     & 2.18    & 2+2      & 2.33  & 1+1     & 1.80        \\ 
    TokenBucket3         & 7+227     & TO        & 4+126   & TO       & 4+20    & 4.16       & 3+68     & 3.16    & 2+2      & 2.18  & 1+1     & 1.62        \\ \hline
\end{tabular}
\caption{Results of \vmtlc using different learners. Columns
    represent: benchmark name (Benchmark), and for each learner's
    contextual and modular approach: the sum of iterations (E+I) and
    execution time in minutes (Time). The `E' value represents the total iterations between
    learner and test generator (i.e. external loop in
    Algorithm~\ref{alg:main}), and `I' represents the total iterations
    the learner took (i.e.\@ loop of Algorithm~\ref{alg:hice}). 
    For the LLM learner, we report the maximum iteration
    counts over two runs, or the counts from the single successful run
    if only one completed. Entries for benchmarks with TO in
    the `T(m)' represent a time out, where the entry in `E+I' column is the number of
    iterations completed before time out.}
    \label{tab:adt_group1}
\end{table}

\subsection*{RQ1: How effective are the proposed techniques at verifying clients modulo tested library contracts, both modular and contextual contracts?}
To answer this question we evaluate our proposed techniques on our benchmarks suite. We present the detailed results of our evaluation in Table~\ref{tab:adt_group1}. We report the results grouped by the choice of ICE learner: \hice, \llm, or \hllm. 
We consider a benchmark to be solved by a technique if it is able to generate contracts for libraries that pass the test generator, and the client is verified to be correct using these contracts. \revision{We run the pure LLM-based pipeline thrice to account for nondeterministic behavior of LLMs.} We gave a timeout of 1000 seconds for the ICE learners to synthesize a solution and \afl was provided with a timeout of 500 seconds, for each round they are called. We ran our tool on an Intel(R) Xeon(R) W-2295 CPU @ 3.00\,GHz with 256\,GB RAM, running Ubuntu 20.04.

For modular contracts, out of 43 benchmarks, \hice solved 19 benchmarks, \hllm solved 25, and the \llm solution solved 38.  
For contextual contracts, \hice solved 21, \hllm solved 28, and \llm solved 41 benchmarks. 
We find our framework fairly effective in solving VMTLC, given that the benchmarks cannot be verified by state-of-the-art automated tools.

Upon further analysis we observed that the failures of \hice and \hllm could be explained significantly by the absence of required predicates in their templates. For example, in the benchmark  \texttt{TokenBucket} a library method contract required an expression over three variables. The pure \llm learner was able to synthesize the expression, but the expression was not covered by the grammar for the \hice learner (octagonal constraints over two variables), and it was not part of the initial set of predicates given by the LLM for the \hllm learner. We leave the improvement of the latter two learners to address these failure modes to future work. 


\revision{
One failure mode we observed for VMTLC is the lack of sufficiently expressive observer methods. For example, in the AtomicLinkedList2 benchmark, the client inserts keys into a linked list and pops them in LIFO order. The assertion asks to show that the last element popped is the first one that was inserted. This can of course be established easily using an observer method that tracks the last element in the list, but there was no such method provided by the library. Consequently, our tool found no modular or contextual contracts that verified the client. We leave the exploration of synthesizing observer methods to future work.

}

\revision{\noindent\paragraph{\bf A small mutant study:}
We study the performance of our framework in rejecting mutants of our benchmark programs that are \emph{incorrect} but pass testing. We systematically create a class of complex incorrect mutants that evade testing (see Appendix~\ref{sec:app:mutation:study} for details). We do this because our clients formulate rich functional specifications that depend on the libraries, which results in simple mutants generated by off-the-shelf mutation generators being found incorrect by the tester. 

We create 30 mutants by applying 2 intentionally unsafe mutation templates on 15 benchmarks that were previously VMTLC-verified using contextual or modular contract synthesis. These templates modify the client by introducing a probabilistically rare execution path inside the program's non-deterministic loop. When triggered, this rare path injects unexpected values (such as inserting a negative number when the client asserts that all elements are positive) altering the local or library state and eventually driving the program to an assertion violation. We observe that the tester is not able to find the bug in any of them. Using the pure LLM Learner, we observe that VMTLC with modular contracts does not prove any of them, i.e., rejects them as expected. VMTLC with contextual contracts verifies 11/30 benchmarks, synthesizing spurious contracts that were not caught by testing due to hard-to-reach paths. See  discussion in Section~\ref{subsec:VMTLCValue} on why we expect contextual contracts to be more likely to be spurious than modular ones. The buggy programs obtained by our mutations are not very natural/realistic, and this study should not be taken as a comprehensive evaluation of VMTLC on incorrect programs.
Our findings suggest, however, that VMTLC fails on many incorrect programs that testing is unable to find fault with (see Section~\ref{subsec:VMTLCValue}).

}

\subsection*{RQ2: Are contextual contracts simpler than modular contracts, and are they easier to synthesize?}


Synthesizing contextual contracts was indeed much more effective (i.e., more benchmarks solved) than synthesizing modular contracts to establish VMTLC. All techniques took less time to synthesize contextual contracts. 
A manual inspection revealed that the synthesized contextual contracts were consistently smaller and less complex than their modular counterparts across all benchmarks.


\subsection*{RQ3: Are the synthesized library method contracts valid?}

Determining the accuracy of synthesized contracts automatically is a nontrivial problem. One can try to ascertain their correctness using a program verification engine, but our benchmarks cannot be proved using such engines (we tried using SeaHorn~\cite{seahorn} and it failed, as mentioned earlier). It is in fact this hardness that motivated us to propose the VMTLC problem in the first place!

\revisiontwo{We first manually vetted all contracts generated by all learners to establish a verified baseline. However, due to the inherent non-determinism of LLMs, subsequent runs might yield syntactically different contracts. To account for this variance, we executed the LLM pipeline two additional times. For these subsequent runs, we utilized a constraint solver to verify whether the newly synthesized contracts were logically implied by our manually vetted baseline. Contracts satisfying this implication were automatically deemed correct, while any that did not were vetted manually.} Our examination found that \emph{all}  synthesized contextual and modular contracts using all learners are correct.

\revision{
The level of guarantee that VMTLC contracts provide is influenced by the quality of the test generator. We therefore investigate the robustness of our results in two ways. First, we check all the final specifications generated by our tool (across learner variations) against the test generator (AFL++) \revisiontwo{for an additional 30 minutes after each run}. 
We found no violations of any synthesized specifications. 
Second, we report on the quality of tests using standard fuzzing metrics tracked by AFL++. 
On average, each of the final contracts were tested against 34 test cases (after AFL++'s removal of cases it deems redundant). The average bitmap coverage, defined as the percentage of edges exercised in the test object, was 50\%, indicating substantial exploration of program behaviors. Finally, the stability metric was 100\%, meaning that executions were fully deterministic and the generated tests were reproducible across runs. 


}

\subsection*{RQ4: How important is it to use a \emph{generalizing} CHC solver to solve VMTLC?}

As argued in the introduction, our VMTCL framework is best served by a generalizing CHC engine, and the learners we implement are generalizing. To evaluate how a non-generalizing CHC solver would fare, we replaced the CHC solver in our tool with CVC5~\cite{cvc5}. With CVC5 as the solver, the tool produced only 10 modular contracts and 15 contextual contracts on our benchmarks, showing a significant reduction in succeeding at VMTLC compared to learning-based CHC solvers.

\section{Related Work}
\label{sec:relwork}

\paragraph{Data-driven inference of library preconditions and invariants.}
Several works consider the problem of synthesizing testable preconditions for
library methods w.r.t.\@ a postcondition
\cite{SankaranarayananCIG08,GehrDV15,PadhiSM16,astorgaprecond,astorgacontract}.
In particular \cite{astorgaprecond,astorgacontract} synthesize weakest
preconditions (w.r.t.\@ not causing exceptions) and a strongest
post-condition corresponding to this precondition, in a similar
setting to ours, where the conditions are phrased in terms of observer
methods of the library.
However, these techniques are not directly applicable in our setting as
the client program may typically need a stronger precondition and a
stronger postcondition to prove its assertions.
\cite{MiltnerPMW20} infers data-structure invariants that are
sufficient to prove a given specification for a library module. The candidate
invariants are tested for inductiveness and sufficiency w.r.t.\@ the
specification. In contrast, we only have access to the data structure via the observer methods, 
and we have a client that needs to be formally proved.

\paragraph{Data-driven inference of library specifications.}
Zhou et al.\@ \cite{ZhouDDJ21} look at a closely related problem,
where they synthesize contracts for library methods used by a client
functional program, using observer methods (or what they call ``method
predicates''), that are adequate to both prove a property of the client and
pass a tester. The main technique they use is an adaptation of 
multi-abduction \cite{ADG2016} to a setting of richer types.
However, there \revision{are} a couple of crucial differences between their setting and ours.
Firstly, they do not infer inductive invariants for recursive client programs: the user is required to supply these along with the property assertion. 
As a consequence, in the resulting VCs the only unknown predicates to be synthesized are the library method contracts, making the problem amenable
to abduction-based techniques.
In contrast, in our setting we need to synthesize loop invariants
in addition to the library method contracts.
In our experience, abduction-based techniques in this setting
(for example \cite{prabhu2021specification}) don't perform well on
this problem.
Not having to synthesize loop invariants also means that data-driven
approaches in the functional setting can get by 
with just positive and negative examples, while in our setting we need to handle Horn-implication examples.
Finally, in contrast to the functional setting, reasoning about heap-manipulating programs and libraries with persistent state is harder and requires more 
sophisticated program logics.

\paragraph{Satisfiability/Synthesis modulo oracles.}
The tester in our setting can be viewed as an oracle that is part of
the synthesis specification. Such oracle-based
synthesis has been 
considered in other work (e.g.\@ \cite{PolgreenRS22}).
However it is not clear if these techniques generalize the examples
given by the oracle.
\cite{PolgreenRS22,MuduliR22} consider the related problem where
given a constraint with unknown black-box functions (whose implementation alone is
available) one tries to find a
satisfying assignment to free variables in the formula that is
consistent with the behaviour of the black-box functions.
While our problem does not exactly fit in this setting, these
techniques could be
used to tackle the complementary problem of falsifying the assertions
in the client program in a way that is consistent with the library
method implementations.
Extending this line of work \cite{MuduliPR24} tries to prove F*
programs correct modulo contracts for library
methods which are tested for consistency with the library
implementation, while \cite{LahiriR22} learns almost-correct
invariants for loops containing opaque calls to library functions.
In contrast to our setting, the contracts are not about
stateful methods, and are either supplied
by the user or not synthesized at all. 

\paragraph{Using LLMs for synthesizing invariants.}
Several researchers have explored the use of LLMs for reasoning about
program verification tasks like precondition and loop
invariant synthesis
\cite{PeiBSSY23,KamathMSCDLLRRS24,kamath2023finding,cao2025clause2inv,wen2024enchanting,king2025llm,MohammedLR0025},
demonstrating their usefulness in coming up with candidate
specifications, and their performance being comparable to
state-of-the-art symbolic techniques when combined with traditional
techniques like static analysis, solvers, and fuzzing.
Our work explores their use in synthesizing library specifications and
in constructing an effective generalizing CHC Solver.

\revision{\paragraph{Assume-Guarantee Testing.} VMTLC has similarities to assume-guarantee testing~\cite{BlundellGP06}, which proposes  decomposing global specifications to modular specifications, in order to have more effective unit testing and predictive testing. However, our work aims for synthesizing modular specifications and have some components (the client) verified while others are only tested.}
\section{Conclusions}
\label{sec:conclusions}
We have presented a significant advance in the problem of verification
of clients using synthesized contracts for libraries, where libraries
are only \emph{tested\,} for validity. In particular, in this work we
have developed a general and novel framework that utilizes
\emph{generalizing CHC solvers}, and techniques for realizing such
solvers using ICE-learning algorithms.
We have proposed and evaluated our framework using several ICE
learning algorithms ranging from those that learn decision-trees,
those that use LLMs, and their combinations 
Our evaluation shows that our technique shows
promise in effectively solving verification modulo tested library
contracts, in settings where clients are small and libraries are
large, and where current automated techniques fail.

Verification modulo testing library contracts presents a new point in
the design space for automated verification---to get scalability to
larger code bases we give up verification guarantees for parts of the
code. A natural question that arises is: \emph{how valuable is this
partial verification given that contracts for libraries are tested,
but not verified}? First, if synthesized contracts for libraries
happen to be valid, there is clear value in the approach, as we have
formally verified the client. Consequently, it is very important to
have effective testing engines to test contracts. As borne out in our
evaluation, an effective testing strategy practically eliminates
invalid contracts, though not provably so. We believe that the chances
that the synthesizer comes up with invalid contracts that pass the
tester \emph{as well as simultaneously} happen to prove a nontrivial
client program correct would be rare. However, much larger scale
experiments are needed to validate this hypothesis in practical
contexts.
We believe our results show that VMTLC has the potential to
make automated verification significantly more applicable to
larger programs than the current state of the art. 

There are several future directions to consider. First, it would be
interesting to evaluate what the guarantee of verification modulo
tested libraries brings over simply testing a program, in practice.
An evaluation on a large corpus of programs, in a particular context,
as to how many correct programs are VMTLC verifiable, and also how
many wrong programs are rejected by the VMTLC framework but not by testing, would be interesting.

On the technical front, we have primarily explored verification of
clients for simple assertions. It would be interesting to extend our
framework to more complex logics, such as clients that manipulate
their own heaps  (in addition to heaps maintained by libraries), and
handle specification logics for clients that involve heaps. The
verification conditions may not lie in decidable logics unlike the
setting of this paper, and the ability to generate counterexamples may
be more limited.


Finally, contextual contracts that stem from our work seem interesting
even under non-automated settings, where verification engineers prove
clients using human written annotations, but with contextual contracts
only tested. This is an appealing idea that is worth exploring
further.

\medskip
\noindent {\bf Acknowledgments:} This work was supported in part by a research grant from Amazon.

\newpage
\section*{Data-Availability Statement}
The artifact supporting this paper, which includes the implementation of our tool \textsc{Dualis}, the complete suite of benchmarks, and the evaluation scripts required to reproduce the experimental results presented in Section 7, is publicly available. To ensure long-term archival, and reproducing the results of this article, the artifact has been packaged and published on Zenodo at \url{https://doi.org/10.5281/zenodo.19519743}. For building upon the framework, the public repository is hosted on GitHub at \url{https://github.com/omar826/Dualis-VMTLC}.
\bibliographystyle{ACM-Reference-Format}
\bibliography{refs}

@inproceedings{houdini,
author = {Flanagan, Cormac and Leino, K. Rustan M.},
title = {Houdini, an Annotation Assistant for ESC/Java},
year = {2001},
isbn = {3540417915},
publisher = {Springer-Verlag},
booktitle = {Proc.\@ Intl.\@ Symp.\@ Formal Methods Europe},
pages = {500–517},
numpages = {18},
series = {FME '01}
}

@InProceedings{spacer1,
author="Komuravelli, Anvesh
and Gurfinkel, Arie
and Chaki, Sagar",
title="SMT-Based Model Checking for Recursive Programs",
booktitle="Proc.\@ 26th Intl.\@ Conf.\@ Computer Aided Verification (CAV 2014), Vienna, Austria, July 18-22, 2014",
year="2014",
publisher="Springer",
pages="17--34",
isbn="978-3-319-08867-9"
}

@inproceedings{spacer2,
author = {Komuravelli, Anvesh and Bj\o{}rner, Nikolaj and Gurfinkel, Arie and McMillan, Kenneth L.},
title = {Compositional verification of procedural programs using horn clauses over integers and arrays},
year = {2015},
isbn = {9780983567851},
publisher = {FMCAD Inc},
address = {Austin, Texas},
booktitle = {Proc.\@ 15th Conf.\@ Formal Methods in Computer-Aided Design},
pages = {89–96},
numpages = {8},
location = {Austin, Texas},
series = {FMCAD '15}
}

@string{LNCS = "LNCS"}

@string{SV = "Springer"}

@article{astorgacontract,
author = {Astorga, Angello and Saha, Shambwaditya and Dinkins, Ahmad and Wang, Felicia and Madhusudan, P. and Xie, Tao},
title = {Synthesizing contracts correct modulo a test generator},
year = {2021},
issue_date = {October 2021},
publisher = {Association for Computing Machinery},
address = {New York, NY, USA},
volume = {5},
number = {OOPSLA},
journal = {Proc. ACM Program. Lang.},
month = oct,
articleno = {104},
numpages = {27}
}

@inproceedings{astorgaprecond,
author = {Astorga, Angello and Madhusudan, P. and Saha, Shambwaditya and Wang, Shiyu and Xie, Tao},
title = {Learning stateful preconditions modulo a test generator},
year = {2019},
isbn = {9781450367127},
publisher = {ACM},
address = {New York, NY, USA},
pages = {775–787},
numpages = {13},
location = {Phoenix, AZ, USA},
series = {PLDI 2019}
}

@inproceedings{icedt,
  author       = {Pranav Garg and
                  Daniel Neider and
                  P. Madhusudan and
                  Dan Roth},
  editor       = {Rastislav Bod{\'{\i}}k and
                  Rupak Majumdar},
  title        = {Learning invariants using decision trees and implication counterexamples},
  booktitle    = {Proc.\@ 43rd {ACM} {SIGPLAN-SIGACT} Symp.\@
                  Principles of Programming Languages ({POPL} 2016), St. Petersburg,
                  USA, 2016},
  pages        = {499--512},
  publisher    = {{ACM}},
  year         = {2016},
  bibsource    = {dblp computer science bibliography, https://dblp.org}
}

@inproceedings{ice,
  author       = {Pranav Garg and
                  Christof L{\"{o}}ding and
                  P. Madhusudan and
                  Daniel Neider},
  editor       = {Armin Biere and
                  Roderick Bloem},
  title        = {{ICE:} {A} Robust Framework for Learning Invariants},
  booktitle    = {Proc.\@ 26th International Conference on Computer Aided Verification {CAV} 2014, Vienna,
                  Austria, 2014.},
  series       = {LNCS},
  volume       = {8559},
  pages        = {69--87},
  publisher    = {Springer},
  year         = {2014},
  bibsource    = {dblp computer science bibliography, https://dblp.org}
}

@inproceedings{Z3,
  author    = {De Moura, Leonardo and Bj{\o}rner, Nikolaj},
  title     = {Z3: An Efficient SMT Solver},
  booktitle = {Proc.\@ 14th Intl.\@ Conf.\@ Tools and Algorithms for the Construction and Analysis of Systems (TACAS 2008)},
  year      = {2008},
  isbn      = {3-540-78799-2, 978-3-540-78799-0},
  location  = {Budapest, Hungary},
  pages     = {337--340},
  numpages  = {4},
  publisher = {Springer-Verlag},
  address   = {Berlin, Heidelberg}
}

@InProceedings{rusthorn,
author="Matsushita, Yusuke
and Tsukada, Takeshi
and Kobayashi, Naoki",
editor="M{\"u}ller, Peter",
title="RustHorn: CHC-Based Verification for Rust Programs",
booktitle="Programming Languages and Systems",
year="2020",
publisher="Springer International Publishing",
address="Cham",
pages="484--514",
isbn="978-3-030-44914-8"
}

@InProceedings{seahorn,
author="Gurfinkel, Arie
and Kahsai, Temesghen
and Komuravelli, Anvesh
and Navas, Jorge A.",
title="The SeaHorn Verification Framework",
booktitle    = {Proc.\@ 27th Intl.\@ Conf.\@ Computer Aided Verification ({CAV} 2015), San Francisco, USA, July 18-24, 2015},
year="2015",
publisher="Springer",
pages="343--361",
isbn="978-3-319-21690-4"
}

@article{DBLP:journals/pacmpl/EzudheenND0M18,
  author    = {P. Ezudheen and
               Daniel Neider and
               Deepak D'Souza and
               Pranav Garg and
               P. Madhusudan},
  title     = {Horn-{ICE} learning for synthesizing invariants and contracts},
  journal   = {{PACMPL}},
  volume    = {2},
  number    = {{OOPSLA}},
  pages     = {131:1--131:25},
  year      = {2018},
}

@inproceedings{KahsaiRSS16,
  author    = {Temesghen Kahsai and
               Philipp R{\"{u}}mmer and
               Huascar Sanchez and
               Martin Sch{\"{a}}f},
  title     = {JayHorn: {A} Framework for Verifying {Java} programs},
  booktitle = {CAV, Part {I}},
  pages     = {352--358},
  year      = {2016},
  volume    = {9779},
  series    = LNCS,
  publisher = SV,
}

@article{kamath2023finding,
  title={Finding inductive loop invariants using large language models},
  author={Kamath, Adharsh and Senthilnathan, Aditya and Chakraborty, Saikat and Deligiannis, Pantazis and Lahiri, Shuvendu K and Lal, Akash and Rastogi, Aseem and Roy, Subhajit and Sharma, Rahul},
  journal={arXiv preprint arXiv:2311.07948},
  year={2023}
}

@inproceedings{KamathMSCDLLRRS24,
  author       = {Adharsh Kamath and
                  Nausheen Mohammed and
                  Aditya Senthilnathan and
                  Saikat Chakraborty and
                  Pantazis Deligiannis and
                  Shuvendu K. Lahiri and
                  Akash Lal and
                  Aseem Rastogi and
                  Subhajit Roy and
                  Rahul Sharma},
  editor       = {Nina Narodytska and
                  Philipp R{\"{u}}mmer},
  title        = {Leveraging LLMs for Program Verification},
  booktitle    = {Proc.\@ Formal Methods in Computer-Aided Design ({FMCAD} 2024), Prague, Czech
                  Republic, October 15-18, 2024},
  pages        = {107--118},
  publisher    = {{IEEE}},
  year         = {2024},
  bibsource    = {dblp computer science bibliography, https://dblp.org}
}

@article{cao2025clause2inv,
  title={Clause2Inv: A Generate-Combine-Check Framework for Loop Invariant Inference},
  author={Cao, Weining and Wu, Guangyuan and Xu, Tangzhi and Yao, Yuan and Wei, Hengfeng and Chen, Taolue and Ma, Xiaoxing},
  journal={Proc.\@ ACM on Software Engineering},
  volume={2},
  number={ISSTA},
  pages={1009--1030},
  year={2025},
  publisher={ACM New York, NY, USA}
}

@inproceedings{wen2024enchanting,
  title={Enchanting program specification synthesis by large language models using static analysis and program verification},
  author={Wen, Cheng and Cao, Jialun and Su, Jie and Xu, Zhiwu and Qin, Shengchao and He, Mengda and Li, Haokun and Cheung, Shing-Chi and Tian, Cong},
  booktitle={Proc.\@ Intl.\@ Conf.\@ Computer Aided Verification (CAV 2024)},
  pages={302--328},
  year={2024},
  organization={Springer}
}

@article{team2024gemini,
  title={Gemini 1.5: Unlocking multimodal understanding across millions of tokens of context},
  author={Team, Gemini and Georgiev, Petko and Lei, Ving Ian and Burnell, Ryan and Bai, Libin and Gulati, Anmol and Tanzer, Garrett and Vincent, Damien and Pan, Zhufeng and Wang, Shibo and others},
  journal={arXiv preprint arXiv:2403.05530},
  year={2024}
}

@article{achiam2023gpt,
  title={{GPT-4} Technical Report},
  author={Achiam, Josh and Adler, Steven and Agarwal, Sandhini and Ahmad, Lama and Akkaya, Ilge and Aleman, Florencia Leoni and Almeida, Diogo and Altenschmidt, Janko and Altman, Sam and Anadkat, Shyamal and others},
  journal={arXiv preprint arXiv:2303.08774},
  year={2023}
}

@inproceedings{lpar/KahsaiKRS17,
  author    = {Temesghen Kahsai and
               Rody Kersten and
               Philipp R{\"{u}}mmer and
               Martin Sch{\"{a}}f},
  title     = {Quantified Heap Invariants for Object-Oriented Programs},
  booktitle = {LPAR},
  series    = {EPiC Series in Computing},
  volume    = {46},
  pages     = {368--384},
  publisher = {EasyChair},
  year      = {2017},
}

@inproceedings{tacas/ChampionC0S18,
  author    = {Adrien Champion and
               Tomoya Chiba and
               Naoki Kobayashi and
               Ryosuke Sato},
  title     = {ICE-Based Refinement Type Discovery for Higher-Order Functional Programs},
  booktitle = {TACAS, Part {I}},
  pages     = {365--384},
  series    = LNCS,
  volume    = {10805},
  publisher = SV,
  year      = {2018},
}

@inproceedings{prabhu2021specification,
  title={{Specification Synthesis with Constrained Horn Clauses}},
  author={Prabhu, Sumanth and Fedyukovich, Grigory and Madhukar, Kumar and D'Souza, Deepak},
  booktitle = {{PLDI}},
  pages     = {1203--1217},
  year={2021},
  organization={ACM}
}

@inproceedings{cvc5,
  author    = {Haniel Barbosa and
               Clark W. Barrett and
               Martin Brain and
               Gereon Kremer and
               Hanna Lachnitt and
               Makai Mann and
               Abdalrhman Mohamed and
               Mudathir Mohamed and
               Aina Niemetz and
               Andres N{\"{o}}tzli and
               Alex Ozdemir and
               Mathias Preiner and
               Andrew Reynolds and
               Ying Sheng and
               Cesare Tinelli and
               Yoni Zohar},
  title     = {cvc5: {A} Versatile and Industrial-Strength {SMT} Solver},
  booktitle = {Proc.\@ 28th Intl.\@ Conf.\@ Tools and Algorithms for the Construction and Analysis of Systems (TACAS 2022), Munich, Germany, April 2-7, 2022},
  series    = {LNCS},
  volume    = {13243},
  pages     = {415--442},
  publisher = {Springer},
  year      = {2022},
  bibsource = {dblp computer science bibliography, https://dblp.org},
}

@inproceedings{arrfreq,
  author    = {Grigory Fedyukovich and Sumanth Prabhu and
               Kumar Madhukar and
               Aarti Gupta},
  title     = {{Quantified Invariants via Syntax-Guided Synthesis}},
  booktitle = {CAV, Part I},
  year      = {2019},
  volume    = {11561},
  pages     = {259--277},
  series    = LNCS,
  publisher = SV,
}

@inproceedings{king2025llm,
  title={Llm-Based Generation of Weakest Preconditions and Precise Array Invariants},
  author={King, Daragh and Koutavas, Vasileios and Kov{\'a}cs, Laura},
  booktitle={Proc.\@ 13th IEEE/ACM Intl.\@ Conf.\@ Formal Methods in Software Engineering (FormaliSE)},
  pages={1--5},
  year={2025},
  organization={IEEE}
}

@book{alur2013syntax,
  title={Syntax-guided synthesis},
  author={Alur, Rajeev and Bodik, Rastislav and Juniwal, Garvit and Martin, Milo MK and Raghothaman, Mukund and Seshia, Sanjit A and Singh, Rishabh and Solar-Lezama, Armando and Torlak, Emina and Udupa, Abhishek},
  year={2013},
  publisher={IEEE}
}

@article{AptOlderogHoareLogic,
author = {Apt, Krzysztof R. and Olderog, Ernst-R\"{u}diger},
title = {Fifty years of Hoare’s logic, Chapter 9},
year = {2019},
issue_date = {Dec 2019},
publisher = {Springer-Verlag},
address = {Berlin, Heidelberg},
volume = {31},
number = {6},
issn = {0934-5043},
journal = {Form. Asp. Comput.},
}

@inproceedings{pldi/GrebenshchikovLPR12,
  author    = {Sergey Grebenshchikov and
               Nuno P. Lopes and
               Corneliu Popeea and
               Andrey Rybalchenko},
  title     = {Synthesizing software verifiers from proof rules},
  booktitle = {PLDI},
  pages     = {405--416},
  year      = {2012},
  publisher = {{ACM}},
}

@inproceedings{pldi/ZhuMJ18,
  author    = {He Zhu and
               Stephen Magill and
               Suresh Jagannathan},
  title     = {A data-driven {CHC} solver},
  booktitle = {PLDI},
  pages     = {707--721},
  year      = {2018},
  publisher = {{ACM}},
}

@inproceedings{Eldarica,
  author    = {Hossein Hojjat and
               Philipp R{\"{u}}mmer},
  title     = {The {ELDARICA} {Horn Solver}},
  booktitle = {{FMCAD}},
  pages     = {158--164},
  publisher = {{IEEE}},
  year      = {2018}
}

@inproceedings{multifreq,
  author    = {Grigory Fedyukovich and Sumanth Prabhu and
               Kumar Madhukar and
               Aarti Gupta},
  title     = {{Solving Constrained Horn Clauses Using Syntax and Data}},
  booktitle = {FMCAD},
  year      = {2018},
  publisher = {IEEE},
  pages     = {170--178},
}

@inproceedings{prabhu2018efficiently,
  title={Efficiently learning safety proofs from appearance as well as behaviours},
  author={Prabhu, Sumanth and Madhukar, Kumar and Venkatesh, R},
  booktitle={SAS},
  pages={326--343},
  year={2018},
  publisher=SV,
  series    = LNCS,
  volume    = {11002},
}

@article{ZhouDDJ21,
  author       = {Zhe Zhou and
                  Robert Dickerson and
                  Benjamin Delaware and
                  Suresh Jagannathan},
  title        = {Data-driven abductive inference of library specifications},
  journal      = {Proc. {ACM} Program. Lang.},
  volume       = {5},
  number       = {{OOPSLA}},
  pages        = {1--29},
  year         = {2021},
  bibsource    = {dblp computer science bibliography, https://dblp.org}
}

@inproceedings{ADG2016,
  author       = {Aws Albarghouthi and
                  Isil Dillig and
                  Arie Gurfinkel},
  title        = {Maximal specification synthesis},
  booktitle    = {Proc.\@ 43rd Annual {ACM} {SIGPLAN-SIGACT} Symp.\@
                  Principles of Programming Languages ({POPL} 2016), St. Petersburg,
                  USA, January 20-22, 2016},
  pages        = {789--801},
  publisher    = {{ACM}},
  year         = {2016},
  bibsource    = {dblp computer science bibliography, https://dblp.org}
}

@inproceedings{beyer2025improvements,
  title={Improvements in software verification and witness validation: SV-COMP 2025},
  author={Beyer, Dirk and Strej{\v{c}}ek, Jan},
  booktitle={International Conference on Tools and Algorithms for the Construction and Analysis of Systems},
  pages={151--186},
  year={2025},
  organization={Springer}
}

@inproceedings{MuduliPR24,
  author       = {Sujit Kumar Muduli and
                  Rohan Ravikumar Padulkar and
                  Subhajit Roy},
  editor       = {Arie Gurfinkel and
                  Vijay Ganesh},
  title        = {Interactive Theorem Proving Modulo Fuzzing},
  booktitle    = {Proc.\@ 36th Intl.\@ Conf.\@ Computer Aided Verification ({CAV} 2024), Montreal, QC, Canada, July 24-27, 2024},
  series       = {LNCS},
  volume       = {14681},
  pages        = {480--493},
  publisher    = {Springer},
  year         = {2024},
  bibsource    = {dblp computer science bibliography, https://dblp.org}
}

@article{MuduliR22,
  author       = {Sujit Kumar Muduli and
                  Subhajit Roy},
  title        = {Satisfiability modulo fuzzing: a synergistic combination of {SMT}
                  solving and fuzzing},
  journal      = {Proc. {ACM} Program. Lang.},
  volume       = {6},
  number       = {{OOPSLA2}},
  pages        = {1236--1263},
  year         = {2022},
  bibsource    = {dblp computer science bibliography, https://dblp.org}
}

@inproceedings{PolgreenRS22,
  author       = {Elizabeth Polgreen and
                  Andrew Reynolds and
                  Sanjit A. Seshia},
  title        = {Satisfiability and Synthesis Modulo Oracles},
  booktitle    = {Proc.\@ 23rd Intl.\@ Conf.\@ Verification, Model Checking, and Abstract Interpretation ({VMCAI} 2022), Philadelphia, USA, January 16-18, 2022},
  series       = {LNCS},
  volume       = {13182},
  pages        = {263--284},
  publisher    = {Springer},
  year         = {2022},
  bibsource    = {dblp computer science bibliography, https://dblp.org}
}

@inproceedings{MiltnerPMW20,
  author       = {Anders Miltner and
                  Saswat Padhi and
                  Todd D. Millstein and
                  David Walker},
  title        = {Data-driven inference of representation invariants},
  booktitle    = {Proc.\@ 41st {ACM} {SIGPLAN} Intl,\@ Conf.\@ Programming Language Design and Implementation ({PLDI} 2020), London,
                  UK, June 15-20, 2020},
  pages        = {1--15},
  publisher    = {{ACM}},
  year         = {2020},
  bibsource    = {dblp computer science bibliography, https://dblp.org}
}

@inproceedings{PadhiSM16,
  author       = {Saswat Padhi and
                  Rahul Sharma and
                  Todd D. Millstein},
  editor       = {Chandra Krintz and
                  Emery D. Berger},
  title        = {Data-driven precondition inference with learned features},
  booktitle    = {Proc.\@ 37th {ACM} {SIGPLAN} Conference on Programming Language Design and Implementation ({PLDI} 2016), Santa Barbara, USA, June 13-17, 2016},
  pages        = {42--56},
  publisher    = {{ACM}},
  year         = {2016},
  bibsource    = {dblp computer science bibliography, https://dblp.org}
}

@inproceedings{GehrDV15,
  author       = {Timon Gehr and
                  Dimitar Dimitrov and
                  Martin T. Vechev},
  title        = {Learning Commutativity Specifications},
  booktitle    = {Proc.\@ 27th Intl.\@ Conf.\@ Computer Aided Verification ({CAV} 2015), San Francisco, USA, July 18-24, 2015},
  series       = {LNCS},
  volume       = {9206},
  pages        = {307--323},
  publisher    = {Springer},
  year         = {2015},
  bibsource    = {dblp computer science bibliography, https://dblp.org}
}

@inproceedings{SankaranarayananCIG08,
  author       = {Sriram Sankaranarayanan and
                  Swarat Chaudhuri and
                  Franjo Ivancic and
                  Aarti Gupta},
  title        = {Dynamic inference of likely data preconditions over predicates by
                  tree learning},
  booktitle    = {Proc.\@ {ACM/SIGSOFT} Intl.\@ Symp.\@ on Software
                  Testing and Analysis ({ISSTA} 2008), Seattle, USA, July 20-24,
                  2008},
  pages        = {295--306},
  publisher    = {{ACM}},
  year         = {2008},
  bibsource    = {dblp computer science bibliography, https://dblp.org}
}

@inproceedings{PeiBSSY23,
  author       = {Kexin Pei and
                  David Bieber and
                  Kensen Shi and
                  Charles Sutton and
                  Pengcheng Yin},
  editor       = {Andreas Krause and
                  Emma Brunskill and
                  Kyunghyun Cho and
                  Barbara Engelhardt and
                  Sivan Sabato and
                  Jonathan Scarlett},
  title        = {Can Large Language Models Reason about Program Invariants?},
  booktitle    = {International Conference on Machine Learning, {ICML} 2023, 23-29 July
                  2023, Honolulu, Hawaii, {USA}},
  series       = {Proc.\@ Machine Learning Research},
  volume       = {202},
  pages        = {27496--27520},
  publisher    = {{PMLR}},
  year         = {2023},
  bibsource    = {dblp computer science bibliography, https://dblp.org}
}

@inproceedings{MohammedLR0025,
  author       = {Nausheen Mohammed and
                  Akash Lal and
                  Aseem Rastogi and
                  Rahul Sharma and
                  Subhajit Roy},
  title        = {{LLM} Assistance for Memory Safety},
  booktitle    = {47th {IEEE/ACM} International Conference on Software Engineering,
                  {ICSE} 2025, Ottawa, ON, Canada, April 26 - May 6, 2025},
  pages        = {1717--1728},
  publisher    = {{IEEE}},
  year         = {2025},
  bibsource    = {dblp computer science bibliography, https://dblp.org}
}

@inproceedings{LahiriR22,
  author       = {Sumit Lahiri and
                  Subhajit Roy},
  editor       = {Sukyoung Ryu and
                  Yannis Smaragdakis},
  title        = {Almost correct invariants: synthesizing inductive invariants by fuzzing
                  proofs},
  booktitle    = {Proc.\@ 31st {ACM} {SIGSOFT} Intl.\@ Symp.\@ on Software Testing and Analysis ({ISSTA} 2022), South Korea, July 18 - 22, 2022},
  pages        = {352--364},
  publisher    = {{ACM}},
  year         = {2022},
  bibsource    = {dblp computer science bibliography, https://dblp.org}
}

@inproceedings{McMillan03,
  author       = {Kenneth L. McMillan},
  editor       = {Warren A. Hunt Jr. and
                  Fabio Somenzi},
  title        = {Interpolation and SAT-Based Model Checking},
  booktitle    = {Proc.\@ 15th Intl.\@ Conf.\@ Computer Aided Verification ({CAV} 2003), Boulder, USA, July 8-12, 2003},
  series       = {LNCS},
  volume       = {2725},
  pages        = {1--13},
  publisher    = {Springer},
  year         = {2003},
  bibsource    = {dblp computer science bibliography, https://dblp.org}
}

@inproceedings{SomenziB11,
  author       = {Fabio Somenzi and
                  Aaron R. Bradley},
  editor       = {Per Bjesse and
                  Anna Slobodov{\'{a}}},
  title        = {{IC3:} where monolithic and incremental meet},
  booktitle    = {Proc.\@ Intl.\@ Conf.\@ Formal Methods in Computer-Aided Design ({FMCAD} 2011), Austin, USA, October 30 - November 02, 2011},
  pages        = {3--8},
  publisher    = {{FMCAD} Inc.},
  year         = {2011},
  bibsource    = {dblp computer science bibliography, https://dblp.org}
}

@article{BlundellGP06,
  author       = {Colin Blundell and
                  Dimitra Giannakopoulou and
                  Corina S. Pasareanu},
  title        = {Assume-Guarantee Testing},
  journal      = {{ACM} {SIGSOFT} Softw. Eng. Notes},
  volume       = {31},
  number       = {2},
  year         = {2006},
}

@inproceedings{JMLLogicBartJacobs2001,
author = {Jacobs, Bart and Poll, Erik},
title = {A Logic for the Java Modeling Language JML},
year = {2001},
isbn = {3540418636},
publisher = {Springer-Verlag},
address = {Berlin, Heidelberg},
booktitle = {Proc.\@ 4th International Conference on Fundamental Approaches to Software Engineering},
pages = {284–299},
numpages = {16},
series = {FASE '01}
}

@inproceedings{JMLTutorial,
author = {Leavens, Gary T.},
title = {Tutorial on JML, the Java Modeling Language},
year = {2007},
isbn = {9781595938824},
publisher = {ACM},
address = {New York, NY, USA},
booktitle = {Proc.\@ 22nd IEEE/ACM Intl.\@ Conf.\@ Automated Software Engineering (ASE 2007)},
pages = {573},
numpages = {1},
location = {Atlanta, Georgia, USA},
}

@article{CodeContracts,
author = {Logozzo, Francesco},
title = {Practical specification and verification with code contracts},
year = {2013},
issue_date = {December 2013},
publisher = {Association for Computing Machinery},
address = {New York, NY, USA},
volume = {33},
number = {3},
issn = {1094-3641},
url = {https://doi.org/10.1145/2658982.2534188},
doi = {10.1145/2658982.2534188},
journal = {Ada Lett.},
month = nov,
pages = {7–8},
numpages = {2}
}

\newpage
\appendix

\section{Benchmark Details}
\label{sec:app:benchmark}
\begin{table}[h!]
  \centering
  \begin{small}
  \begin{tabular}{|l|r|r|r|r|r|}
    \hline
    \textbf{Name} & \textbf{Lbry LOC} & \textbf{Clnt LOC} & \textbf{\#Obs-Mthds} & \textbf{\#Mthd Specs} & \textbf{\#LoopInvs} \\ \hline
    AlternatingList      & 100 & 15 & 2 & 1 & 1 \\ \hline
    AtomicHashMap (1 - 5) & $\sim$3k & 17 - 28 & 4 & 1 - 2 & 1 - 3\\ \hline
    AtomicLinkedList (1 - 2)  & $\sim$500 & 16 - 19 & 1 - 2 & 2& 2 \\ \hline
    BinaryHeap (1 - 2)        & $\sim$200 & 15 - 19 & 2 - 3 & 3& 1 - 2 \\ \hline
    BinaryTree           & 100 - 200 & 14 & 2 & 2& 1 \\ \hline
    BlueWhite            & $\sim$100 & 14 & 1 & 1& 2 \\ \hline
    Calendar             & $\sim$100 & 12 & 2 & 1& 1 \\ \hline
    DLL\_Circular        & $\sim$100 & 12 & 1 & 1& 1 \\ \hline
    DLL\_Token           & $\sim$100 & 7  & 2 & 1& 1 \\ \hline
    FlatHashMap (1 - 2)  & $\sim$3k & 16 - 22 & 3 & 2&1 - 2 \\ \hline
    FlatHashSet          & $\sim$3k & 30 & 4 & 2& 3 \\ \hline
    LruCache             & $\sim$3k & 19 & 3 & 2& 1 \\ \hline
    List (Max)                & $\sim$100 & 13 & 2& 1& 1 \\ \hline
    List (Min)                  & $\sim$100 & 13 & 2& 1& 1 \\ \hline
    Multimap (1 - 2)     & $\sim$3k & 12 - 21 & 2 - 3& 1& 1 \\ \hline
    Multiset (1 - 2)     & $\sim$3k & 13 - 20 & 2 - 3& 1& 1 \\ \hline
    NormalFilterQueue    & $\sim$100& 16 & 3& 2& 1\\ \hline
    PriorityFilterQueue  & $\sim$100& 16 & 3& 2& 1\\ \hline
    ProcessQueue         & $\sim$100& 15 & 2& 2& 1\\ \hline
    RedBlackTree         & $\sim$100& 17 & 4& 2& 1\\ \hline
    SkipList (1 - 7)     & $\sim$1000 & 14 - 16 & 4& 2& 1 \\ \hline
    Stack                & $\sim$50 & 18 & 1& 2& 1 \\ \hline
    StockOrder           & $\sim$100 & 13 & 2& 1& 1 \\ \hline
    TokenBucket(1 - 3)   & $\sim$200 & 15 - 18 & 1& 2& 1 \\ \hline
  \end{tabular}
  \end{small}
  \caption{Features of Benchmarks. Columns represent: benchmark name (name), lines of code of library (Lbry LOC),  lines of code of client code (Clnt LOC), number of observer methods in the library (\#Obs-methods), number of methods to synthesize specifications (\#Mthd Specs), and number of loops (\#LoopInvs).}
  \label{tab:benchmarks}
    \end{table}

\revision{\section{\textbf{Mutation Study}}}
\label{sec:app:mutation:study}
\revision{This section presents additional details of the mutation study used to evaluate the effectiveness of VMTLC beyond testing.
}
\newpage
\revision{\subsection{\textbf{Results}}
\begin{table}[h!]
\centering
\begin{small}
\begin{tabular}{|l|c|c|c|}
\hline
\textbf{\shortstack{Mutant\\Benchmarks}} & \textbf{\shortstack{Passed by\\Testing}} & \textbf{\shortstack{Passed by\\Contextual\\VMTLC}} & \textbf{\shortstack{Passed by\\Modular\\VMTLC}} \\ \hline
AtomicHashMap1Mut1 & $\checkmark$ & $\checkmark$ & $\times$ \\ \hline
AtomicHashMap1Mut2 & $\checkmark$ & $\times$ & $\times$ \\ \hline
AtomicLinkedList1Mut1 & $\checkmark$ & $\times$ & $\times$ \\ \hline
AtomicLinkedList1Mut2 & $\checkmark$ & $\times$ & $\times$ \\ \hline
BinaryTreeMut1 & $\checkmark$ & $\checkmark$ & $\times$ \\ \hline
BinaryTreeMut2 & $\checkmark$ & $\checkmark$ & $\times$ \\ \hline
BlueWhiteMut1 & $\checkmark$ & $\times$ & $\times$ \\ \hline
BlueWhiteMut2 & $\checkmark$ & $\checkmark$ & $\times$ \\ \hline
CalendarMut1 & $\checkmark$ & $\times$ & $\times$ \\ \hline
CalendarMut2 & $\checkmark$ & $\times$ & $\times$ \\ \hline
DLL\_CircularMut1 & $\checkmark$ & $\checkmark$ & $\times$ \\ \hline
DLL\_CircularMut2 & $\checkmark$ & $\times$ & $\times$ \\ \hline
FlatHashSetMut1 & $\checkmark$ & $\times$ & $\times$ \\ \hline
FlatHashSetMut2 & $\checkmark$ & $\times$ & $\times$ \\ \hline
MaxMut1 & $\checkmark$ & $\times$ & $\times$ \\ \hline
MaxMut2 & $\checkmark$ & $\times$ & $\times$ \\ \hline
MinMut1 & $\checkmark$ & $\times$ & $\times$ \\ \hline
MinMut2 & $\checkmark$ & $\times$ & $\times$ \\ \hline
Multimap1Mut1 & $\checkmark$ & $\checkmark$ & $\times$ \\ \hline
Multimap1Mut2 & $\checkmark$ & $\times$ & $\times$ \\ \hline
NormalFilterQueueMut1 & $\checkmark$ & $\times$ & $\times$ \\ \hline
NormalFilterQueueMut2 & $\checkmark$ & $\times$ & $\times$ \\ \hline
ProcessQueueMut1 & $\checkmark$ & $\checkmark$ & $\times$ \\ \hline
ProcessQueueMut2 & $\checkmark$ & $\checkmark$ & $\times$ \\ \hline
RedBlackTreeMut1 & $\checkmark$ & $\times$ & $\times$ \\ \hline
RedBlackTreeMut2 & $\checkmark$ & $\times$ & $\times$ \\ \hline
SkipList1Mut1 & $\checkmark$ & $\checkmark$ & $\times$ \\ \hline
SkipList1Mut2 & $\checkmark$ & $\checkmark$ & $\times$ \\ \hline
StackMut1 & $\checkmark$ & $\times$ & $\times$ \\ \hline
StackMut2 & $\checkmark$ & $\checkmark$ & $\times$ \\ \hline
\end{tabular}
\end{small}
\caption{Mutant Benchmarks that passed/failed Testing(30/0), Contextual VMTLC(11/19), and Modular VMTLC(0/30)
}
\end{table}
}
\revision{\subsection{\textbf{Mutation Templates}}}
\revision{To systematically introduce meaningful faults, we designed two mutation templates that inject unsafe behavior through probabilistically rare execution paths.
Both templates are designed to hide unsafe states behind rare conditions that are difficult for testing to explore.
}

\begin{itemize}
\item \revision{\textbf{Mutation Template 1}}
{
\begin{verbatim}
while (*) {
    v=*
    if ( v == Const) {
        count++
    } else {
        count=0
    }
    if (count = 100) {
        rare operation
    } else {
        normal operation
    }
    //other operations
}
assert (client condition)
}
\end{verbatim}
}
\revision{\item \textbf{Mutation Template 2}}
{
\begin{verbatim}
while (*) {
    v=*
    if ( v == Const) {
        count++
    } else {
        count=0
    }
    //other operations
}
if (count = 100) {
    rare operation
} else{
    normal operation
}
assert (client condition)
\end{verbatim}
}
\end{itemize}

\revision{\subsection{Why did 11 mutations pass VMTLC verification for contextual contracts?}
The rare operations in both the above templates are difficult for the tester to explore and hence the learner is able to give any strong adequate contextual contract which the verifier can accept to prove the mutation.
}

\section{LLM Prompts}
\label{sec:app:llmprompts}

\subsection{\textbf{Initial Prompt for Modular Contract Generation}}
\label{sec:app:prompt:init}

\mbox{}\\[0.5em]
You are an expert programmer specializing in formal verification with Z3Py. Your task is to define specifications for general-purpose library functions (e.g., insert, search). You will also be given a client program (in the Abstract and CHC sections) that uses these functions in a specific way. Your goal is to find a Z3Py specification for each library function that meets two conditions:

\begin{enumerate}
  \item \textbf{The specification must encapsulate the library's behavior}, meaning it must be true for any valid inputs the library function might receive, not just the specific inputs used by the client.
  \item \textbf{The specification must be strong enough to prove the assertions of the client program} when used within its specific context.
\end{enumerate}

The ultimate goal is for your generated Z3Py definitions to make all the client's CHC rules valid.

\textbf{I. System Behavior Overview:}

Please analyze the following abstract program description to understand the intended behavior, state variables, and operations. This provides semantic context for the relations you will define.

\begin{center}
\fbox{\texttt{[Client Program Pseudo Code]}}
\end{center}

Z3 integer variables are used to model non-deterministic choices, indicated by \texttt{[*]}.

\textbf{II. Formal CHC Specification Context:}

Please analyze the following CHC content. This is the formal basis for your task.

\begin{enumerate}
  \item \textbf{\texttt{declare-rel} statements:} These define the names and expected arguments (types and order) for the Python/Z3Py functions you must create.
  \item \textbf{\texttt{declare-var} statements:} These list the symbolic variables used within the CHC rules.
  \item \textbf{\texttt{rule} statements:} These are the most critical part. Each rule \texttt{(=> antecedent\_formula consequent\_formula)} defines a logical implication. Your Z3Py function definitions must ensure that \textbf{every one of these rules translates into a VALID Z3 implication}.
  \item \textbf{extra info:} \texttt{ite} stands for if then else.
\end{enumerate}

\begin{center}
\fbox{\texttt{[CHC Rules]}}
\end{center}

\textbf{III. Pre-defined Z3 Variables and Constants in the Testing Environment:}

The Python script where your generated functions will be inserted has ALREADY DEFINED global Z3 variables and Python constants. \textbf{DO NOT redefine these in your output.} The parameters of your functions will be called with these Z3 variables (or expressions derived from them) in the testing script.

\begin{itemize}
  \item \textbf{Globally Available Z3 Integer Variables (These are examples):}
    \begin{itemize}
      \item \texttt{len} (representing current state values for some benchmarks)
      \item \texttt{len1} (representing next state values for some benchmarks)
      \item \texttt{prio\_input} (representing non-deterministic inputs to a loop iteration, if applicable to the benchmark)
    \end{itemize}
  \item \textbf{Globally Available Z3 Boolean Variables (if applicable, based on \texttt{declare-var}):}
    \begin{itemize}
      \item \texttt{ret1} (e.g., for a return status)
    \end{itemize}
  \item \textbf{Globally Available Python Constants (These are examples; refer to \texttt{define-fun} in CHC.txt or initial values in \texttt{Abst\_Program.txt} for benchmark-specific constants):}
    \begin{itemize}
      \item \texttt{INT\_MAX = 32767}
      \item \texttt{INIT\_MIN\_PRIO = 1}
      \item \texttt{INIT\_MAX\_PACKET\_SIZE = 499}
    \end{itemize}
\end{itemize}

\textbf{IV. Your Task: Define Z3Py Function Bodies for Declared Relations}

\begin{enumerate}
  \item \textbf{Identify Target Functions:} For each \texttt{(declare-rel relation\_name (Type1 Type2 ...))} statement in the CHC content (Section II), you must define a corresponding Python function.
  \item \textbf{Function Signatures:}
        These are the function signatures:
    \begin{center}
\fbox{\texttt{[Function Signatures]}}
\end{center}

  \item \textbf{Function Logic (The Core Task):}
    \begin{itemize}
      \item The body of each Python function must construct and return a single Z3 boolean expression.
      \item This Z3 expression \textit{is} the definition of the relation.
      \item The logic for this expression should be derived from your understanding of the \texttt{Abst\_Program.txt} (Section I) and, most importantly, by analyzing how the relation is used within the \textbf{\texttt{rule} statements} of the \texttt{CHC.txt} (Section II). The goal is to define the relation such that all rules involving it become valid implications.
      \item For example, an \texttt{inv\_loop} function should define properties that make the CHC rules for initialization and preservation hold. An \texttt{append\_op} function should define state changes that are consistent with the rules where \texttt{append\_op} appears.
    \end{itemize}
\end{enumerate}

\textbf{V. Expected Output Format:}

Provide \textbf{ONLY} the complete Python code definitions for the required functions. Do not include any surrounding text, explanations, or \texttt{from z3 import *}. Do not redefine any global Z3 variables or constants (as per Section III).

\textbf{Example of Expected Output Structure (for a hypothetical benchmark):}

\begin{verbatim}
def inv_relation_from_chc(p_arg1, p_arg2):
  # Logic using p_arg1, p_arg2, and implicitly relying on how these
  # parameters will be mapped to global Z3 variables in the CHC rules.
  # Constants like INT_MAX can be used if they are listed in Section III.
  return And(p_arg1 > 0, p_arg2 < INT_MAX)

def operation_relation_from_chc(p_curr_state_arg, p_next_state_arg, p_input_arg):
  # Logic defining the transition
  return (p_next_state_arg == p_curr_state_arg + p_input_arg - 1)
\end{verbatim}

\textbf{VI. Additional Guidelines \& Goal Clarification:}

\begin{itemize}
  \item \textbf{Primary Goal: Validity.} The most important objective is that your generated Z3Py function definitions make ALL CHC rules (from Section II), when translated into Z3 implications, \textbf{VALID}.
  \item \textbf{Plausibility:} The library functions you are defining are general-purpose. For example, an insert function can accept any valid integer, even if the client program in the abstract only ever calls it with 1 and 2.
  \item For the key verification conditions (like invariant preservation for different paths), the antecedent of the implication should be satisfiable.
  \item \textbf{Maximal Specification (Secondary Goal, Strive For):}
    \begin{itemize}
      \item Once validity is achieved, aim for your specifications to be "maximal." Think of a specification (like an invariant \texttt{inv(state)} or an operation \texttt{op(current\_state, next\_state)}) as defining a set of allowed states or state transitions.
      \item A valid specification is maximal if you cannot add any more states (for an invariant) or any more input/output state pairs (for an operation) to the set it accepts without making the specification invalid (i.e., without causing one of the CHC rules to no longer hold).
      \item Essentially, try to make your Z3 definitions as general or "permissive" as possible, including all behaviors that are consistent with the CHC rules and the abstract program's intent. Avoid being overly restrictive if a more general valid definition exists.
      \item There might be multiple different maximal specifications. Your goal is to find one that is as comprehensive as possible.
      \item Crucially, prioritize achieving VALIDITY of all CHC rules above striving for maximality if there's a conflict. A valid, slightly less general specification is much better than a more general but invalid one.
    \end{itemize}
  \item Consider edge cases (e.g., empty queues if \texttt{len} is a variable) as implied by the abstract program and CHC rules.
  \item Pay close attention to how variables declared with \texttt{(declare-var ...)} in the CHC rules are used as arguments to your functions within those rules. This mapping is key.
\end{itemize}

\textbf{VII. Tips}

To succeed, you MUST follow this strategic guidance:

\begin{enumerate}
  \item \textbf{Think Relationally, Not Mathematically:} Your primary goal is to define the relationship between the state at one step and the state at the next.
  \item \textbf{Prioritize the General Function First:} Your mental process should be:
    \begin{itemize}
      \item \textbf{First:} Define a plausible, general-purpose library function based on common data structure behavior.
      \item \textbf{Then:} Define the invariant (\texttt{inv1}, \texttt{inv2}) to be a property that is true for the \textit{client's specific usage} of that general function. The invariant proves the client correct; the function defines the library.
    \end{itemize}
  \item \textbf{How to Interpret Counterexamples:} When you receive a counterexample, do not just patch the code to handle that specific case. Instead, try to infer the \textbf{simplest, most general logical rule} that the counterexample reveals.
  \item \textbf{The code is not wrong:} There are no issues from my side. Rethink the logic if you think you are stuck.
  \item \textbf{keep it simple:} Try to find simple contracts that satisfy validity.
\end{enumerate}

\subsection{\textbf{Initial Prompt for contextual contract generation}}
\label{sec:app:prompt:contextual}

\mbox{}\\[0.5em]
You are an expert programmer specializing in formal verification with Z3Py. Your task is to analyze a client program (described in the Abstract and CHC sections) and define Z3Py specifications for the relations it uses. Your goal is to find a specification that is true for the relations within the specific context of the client program's execution flow. The ultimate goal is for these generated contextual specifications to make all the client's CHC rules valid.

\textbf{I. System Behavior Overview:}

Please analyze the following abstract program description to understand the intended behavior, state variables, and operations. This provides semantic context for the relations you will define.

\textit{Insert the abstract program description below:}
\begin{center}
\fbox{\texttt{[Client Program Pseudo Code]}}
\end{center}

Z3 integer variables are used to model non-deterministic choices, indicated by \texttt{[*]}.

\textbf{II. Formal CHC Specification Context:}

Please analyze the following CHC content. This is the formal basis for your task.

\begin{enumerate}
  \item \textbf{\texttt{declare-rel} statements:} These define the names and expected arguments (types and order) for the Python/Z3Py functions you must create.
  \item \textbf{\texttt{declare-var} statements:} These list the symbolic variables used within the CHC rules.
  \item \textbf{\texttt{rule} statements:} These are the most critical part. Each rule \texttt{(=> antecedent\_formula consequent\_formula)} defines a logical implication. Your Z3Py function definitions must ensure that \textbf{every one of these rules translates into a VALID Z3 implication}.
  \item \textbf{extra info:} \texttt{ite} stands for if then else.
\end{enumerate}

\textit{Insert the CHC content below:}
\begin{center}
\fbox{\texttt{[CHC]}}
\end{center}

\textbf{III. Pre-defined Z3 Variables and Constants in the Testing Environment:}

The Python script where your generated functions will be inserted has ALREADY DEFINED global Z3 variables and Python constants. \textbf{DO NOT redefine these in your output.} The parameters of your functions will be called with these Z3 variables (or expressions derived from them) in the testing script.

\begin{itemize}
  \item \textbf{Globally Available Z3 Integer Variables (These are examples):}
    \begin{itemize}
      \item \texttt{len} (representing current state values for some benchmarks)
      \item \texttt{len1} (representing next state values for some benchmarks)
      \item \texttt{prio\_input} (representing non-deterministic inputs to a loop iteration, if applicable to the benchmark)
    \end{itemize}
  \item \textbf{Globally Available Z3 Boolean Variables (if applicable, based on \texttt{declare-var}):}
    \begin{itemize}
      \item \texttt{ret1} (e.g., for a return status)
    \end{itemize}
  \item \textbf{Globally Available Python Constants (These are examples; refer to \texttt{define-fun} in CHC.txt or initial values in \texttt{Abst\_Program.txt} for benchmark-specific constants):}
    \begin{itemize}
      \item \texttt{INT\_MAX = 32767}
      \item \texttt{INIT\_MIN\_PRIO = 1}
      \item \texttt{INIT\_MAX\_PACKET\_SIZE = 499}
    \end{itemize}
\end{itemize}

\textbf{IV. Your Task: Define Z3Py Function Bodies for Declared Relations}

\begin{enumerate}
  \item \textbf{Identify Target Functions:} For each \texttt{(declare-rel relation\_name (Type1 Type2 ...))} statement in the CHC content (Section II), you must define a corresponding Python function.
  \item \textbf{Function Signatures:}
        These are the function signatures. Do not define any other function beyond these.:
    \textit{Insert the function signatures below:}
    \begin{center}
    \fbox{\texttt{[signatures]}}
    \end{center}
  \item \textbf{Function Logic (The Core Task):}
    \begin{itemize}
      \item The body of each Python function must construct and return a single Z3 boolean expression.
      \item This Z3 expression \textit{is} the definition of the relation.
      \item The logic for this expression should be derived from your understanding of the \texttt{Abst\_Program.txt} (Section I) and, most importantly, by analyzing how the relation is used within the \textbf{\texttt{rule} statements} of the \texttt{CHC.txt} (Section II). The goal is to define the relation such that all rules involving it become valid implications.
      \item For example, an \texttt{inv\_loop} function should define properties that make the CHC rules for initialization and preservation hold. An \texttt{append\_op} function should define state changes that are consistent with the rules where \texttt{append\_op} appears.
    \end{itemize}
\end{enumerate}

\textbf{V. Expected Output Format:}

Provide \textbf{ONLY} the complete Python code definitions for the required functions. Do not include any surrounding text, explanations, or \texttt{from z3 import *}. Do not redefine any global Z3 variables or constants (as per Section III).

\textbf{Example of Expected Output Structure (for a hypothetical benchmark):}

\begin{verbatim}
def inv_relation_from_chc(p_arg1, p_arg2):
  # Logic using p_arg1, p_arg2, and implicitly relying on how these
  # parameters will be mapped to global Z3 variables in the CHC rules.
  # Constants like INT_MAX can be used if they are listed in Section III.
  return And(p_arg1 > 0, p_arg2 < INT_MAX)

def operation_relation_from_chc(p_curr_state_arg, p_next_state_arg, p_input_arg):
  # Logic defining the transition
  return (p_next_state_arg == p_curr_state_arg + p_input_arg - 1)
\end{verbatim}

\textbf{VI. Additional Guidelines \& Goal Clarification:}

\begin{itemize}
  \item \textbf{Primary Goal: Validity.} The most important objective is that your generated Z3Py function definitions make ALL CHC rules (from Section II), when translated into Z3 implications, \textbf{VALID}. These function definitions will be tested against their corresponding library functions only on inputs that occur in the execution of the Client Program (Abstract).
  \item For the key verification conditions (like invariant preservation for different paths), the antecedent of the implication should be satisfiable.
  \item Consider edge cases (e.g., empty queues if \texttt{len} is a variable) as implied by the abstract program and CHC rules.
  \item Pay close attention to how variables declared with \texttt{(declare-var ...)} in the CHC rules are used as arguments to your functions within those rules. This mapping is key.
\end{itemize}

\textbf{VII. Tips}

To succeed, you MUST follow this strategic guidance:

\begin{enumerate}
  \item \textbf{Think Relationally, Not Mathematically:} Your primary goal is to define the relationship between the state at one step and the state at the next.
  \item \textbf{How to Interpret Counterexamples:} When you receive a counterexample, do not just patch the code to handle that specific case. Instead, try to infer the \textbf{simplest, most general logical rule} that the counterexample reveals.
  \item \textbf{ The code is not wrong:} There are no issues from my side. Rethink the logic if you think you are stuck.
  \item \textbf{ keep it simple:} Try to find simple contracts that satisfy validity.
\end{enumerate}

\subsection{\textbf{Repair Prompt for Z3 Validity Failures}}
\label{sec:app:prompt:repair1}
\mbox{}\\[0.5em]

The previous Z3 code was invalid. The validator failed with this output:

        \begin{center}
        \fbox{\texttt{[Validity Checker Report]}}
        \end{center}
  
        Your task is to fix the code based on the validator's output.
        Respond with ONLY the complete, corrected, and runnable Python code for all functions.
        Do not include any explanations, apologies, introductory text, or markdown formatting like ```python.
        Your entire response must be valid Python code.
        
\subsection{\textbf{Repair Prompt for Testing Failures}}
\label{sec:app:prompt:repair2}
\mbox{}\\[0.5em]

The previous Z3 specification was valid but failed testing.
The counterexample data is:

\begin{center}
\fbox{\texttt{[Tester Report]}}
\end{center}

Your task is to analyze the counterexample and provide a new, corrected version of the Z3 code to fix the logical flaw.
Respond with ONLY the complete, corrected, and runnable Python code for all functions.
Do not include any explanations, introductory text, or markdown formatting like ```python.
Your entire response must be valid Python code.

\subsection{\textbf{Initial Prompt for \hllm}}
\label{sec:app:prompt:hllm}
\mbox{}\\[0.5em]

You are an expert in program semantics and formal verification. Your primary task is to analyze an abstract program description and its corresponding CHC specification. For each relation declared in the CHC file (like \texttt{inv}, \texttt{insert}, etc.), you will generate a textual description of its parameters and a list of core arithmetic and logical sub-expressions that describe the relationships between those parameters.

\textbf{I. System Behavior Overview:}

Please analyze the following abstract program description to understand the intended behavior, state variables, and operations. This provides semantic context for the relations you will analyze.

\begin{center}
\fbox{\texttt{[Client Program Pseudo Code]}}
\end{center}

\textbf{II. Formal CHC Specification Context:}

Please analyze the following CHC content. This content primarily serves to:

\begin{itemize}
  \item Identify the names and expected arguments (types and order) for the Python functions you must create (from \texttt{declare-rel} statements).
  \item Provide context on how these relations are used in logical rules, which will inform the plausible semantics you need to model.
  \item List the symbolic variables involved (from \texttt{declare-var} statements).
\end{itemize}

\begin{center}
\fbox{\texttt{[CHC Rules]}}
\end{center}

\textbf{III. Pre-defined Variable and Constant Context:}

The CHC file may use pre-defined constants. Your analysis should be consistent with these definitions.

Example: \texttt{(define-fun INT\_MAX () Int 32767)} means \texttt{INT\_MAX} is 32767.

\textbf{IV. Your Task: Generate Textual Expression Sets for Each Relation}

For each relation defined by a \texttt{(declare-rel ...)} statement in Section II, you must:

\begin{itemize}
  \item Analyze its plausible behavior based on its name, its arguments, the abstract program, and its usage in the CHC rules.
  \item Identify the core arithmetic relationships between its parameters. For example, if a relation describes \texttt{len1 = len + 1}, a key expression is \texttt{len1 - len }.
  \item Format your output for that relation exactly as described in Section V below.
\end{itemize}

\textbf{Function Signatures:} These are the function signatures:

\begin{center}
\fbox{\texttt{[Function Signatures]}}
\end{center}

\textbf{Function Logic (The Core Task):}

From the behavior of these functions, you will extract key arithmetic or boolean terms.

\textbf{Canonical Form:} Where possible, express relationships as terms that would be compared to zero. For example, if you infer a behavior \texttt{x = y + 2}, the corresponding expression to return in the list would be \texttt{x - y}. If you infer \texttt{a > b}, a good expression would be \texttt{a - b}.

\textbf{V. Expected Output Format:}

Provide your response as a block of text. For each relation, you must generate a multi-line entry strictly following this format:

\begin{itemize}
  \item A header line: \texttt{<relation\_name> <num\_parameters> <num\_expressions>}
  \item A list of parameter mappings, one per line: \texttt{<index> <parameter\_name>}
  \item A list of the extracted expressions, one per line, using the parameter names you just defined.
\end{itemize}

\textbf{Example of the Required Output Format (for a hypothetical benchmark):}

\begin{verbatim}
insert 5 7
0 n
1 min
2 min1
3 isEmpty
4 isEmpty1
n - min1
min1 - n
n - min
min - n
min1 - min
min - min1
min1 - n

search 4 2
0 v
1 min
2 isEmpty
3 ret1
v - min
min - v
\end{verbatim}

\textbf{VI. Goal Clarification:}

\textbf{Primary Goal: Useful Expressions.} Your goal is to provide a rich set of fundamental expressions (building blocks). A downstream solver will take these expressions \texttt{e} and try to find a valid specification by combining them into predicates of the form \texttt{e > c}, or \texttt{e = c} (where \texttt{c} is a constant). Your generated expressions should be these \texttt{e} terms.

\textbf{Completeness:} Aim to provide a comprehensive set of simple expressions that capture the relationships. For example, if \texttt{a} and \texttt{b} are compared, providing both \texttt{a - b} and \texttt{b - a} can be useful for the solver.

\subsection{\textbf{initial prompt for hornice+llm (contextual)}}
\label{sec:app:prompt:hornice-llm-contextual}

\mbox{}\\[0.5em]
You are an expert in program semantics and formal verification. Your task is to analyze a specific client program and its corresponding CHC specification. For each relation (like inv, insert, etc.), you will generate a list of core arithmetic sub-expressions. These expressions must describe the relationships between the function parameters only as they occur within the specific execution flow of the client program.

\textbf{I. System Behavior Overview:}

Please analyze the following abstract program description to understand the intended behavior, state variables, and operations. This client program description defines the exact execution context for your analysis. The expressions you generate must be consistent with the state changes and variable values that can occur within this specific program.

\begin{center}
\fbox{\texttt{[Client Program Pseudo Code]}}
\end{center}

\textbf{II. Formal CHC Specification Context:}

Please analyze the following CHC content. This content primarily serves to:
\begin{itemize}
    \item Identify the names and expected arguments (types and order) for the Python functions you must create (from \texttt{declare-rel} statements).
    \item Provide context on how these relations are used in logical rules, which will inform the plausible semantics you need to model.
    \item List the symbolic variables involved (from \texttt{declare-var} statements).
\end{itemize}

\begin{center}
\fbox{\texttt{[CHC Rules]}}
\end{center}

\textbf{III. Pre-defined Variable and Constant Context:}

The CHC file may use pre-defined constants. Your analysis should be consistent with these definitions.

\textbf{Example:} \texttt{(define-fun INT\_MAX () Int 32767)} means \texttt{INT\_MAX} is 32767.

\textbf{IV. Your Task: Generate Textual Expression Sets for Each Relation}

For each relation defined by a \texttt{(declare-rel ...)} statement in Section II, you must:

\begin{itemize}
    \item Analyze its plausible behavior based on its name, its arguments, the abstract program, and its usage in the CHC rules.
    \item Identify the core arithmetic relationships between its parameters. For example, if a relation describes \texttt{len1 = len + 1}, a key expression is \texttt{len1 - len}.
    \item Format your output for that relation exactly as described in Section V below.
\end{itemize}

\textbf{Function Signatures:} These are the function signatures:

\begin{center}
\fbox{\texttt{[Function Signatures]}}
\end{center}

\textbf{Function Logic (The Core Task):}
\begin{itemize}
    \item From the behavior of these functions, you will extract key arithmetic or boolean terms.
    \item \textbf{Canonical Form:} Where possible, express relationships as terms that would be compared to zero. For example, if you infer a behavior \texttt{x = y + 2}, the corresponding expression to return in the list would be \texttt{x - y}. If you infer \texttt{a > b}, a good expression would be \texttt{a - b}.
\end{itemize}

\textbf{V. Expected Output Format:}

Provide your response as a block of text. For each relation, you must generate a multi-line entry strictly following this format:
\begin{itemize}
    \item A header line: \texttt{<relation\_name> <num\_parameters> <num\_expressions>}
    \item A list of parameter mappings, one per line: \texttt{<index> <parameter\_name>}
    \item A list of the extracted expressions, one per line, using the parameter names you just defined.
\end{itemize}

\textbf{Example of the Required Output Format (for a hypothetical benchmark):}
\begin{verbatim}
insert 5 7
0 n
1 min
2 min1
3 isEmpty
4 isEmpty1
n - min1
min1 - n
n - min
min - n
min1 - min
min - min1
min1 - n

search 4 2
0 v
1 min
2 isEmpty
3 ret1
v - min
min - v
\end{verbatim}

\textbf{VI. Goal Clarification:}
\begin{itemize}
    \item \textbf{Primary Goal: Useful Expressions.} Your goal is to provide a rich set of fundamental expressions (building blocks). A downstream solver will take these expressions \texttt{e} and try to find a valid specification by combining them into predicates of the form \texttt{e > c}, or \texttt{e = c} (where \texttt{c} is a constant). Your generated expressions should be these \texttt{e} terms.
    \item \textbf{Completeness:} Aim to provide a comprehensive set of simple expressions that capture the relationships. For example, if \texttt{a} and \texttt{b} are compared, providing both \texttt{a - b} and \texttt{b - a} can be useful for the solver.
    \item \textbf{Focus on Context, Not Generality:} The expressions you generate should model the function's behavior only as it's called by the client program. For example, if the client's logic ensures an \texttt{insert} function is only ever called with positive integers, your expressions should reflect this constraint. You do not need to generate expressions for behaviors that are impossible within this specific client's execution flow.
    \item \textbf{Allowed Symbols:} Each expression should be constructed using variable names, and the symbols: \texttt{(}, \texttt{)}, \texttt{+}, \texttt{-}. Integers, logical expressions etc are not allowed.
\end{itemize}


\end{document}